\documentclass[aps,prc,reprint,superscriptaddress]{revtex4-2}

\usepackage{graphicx,subfigure}
\usepackage{dcolumn}
\usepackage{bm}
\usepackage{xcolor}
\usepackage{multirow}
\usepackage[utf8]{inputenc}
\usepackage[T1]{fontenc}

\begin{document}
\preprint{APS/123-QED}

\title{Three-Nucleon Dynamics in the dp breakup  collisions at 190 MeV/nucleon using the WASA detector at COSY-J\"{u}lich} 



\newcommand*{\IKPUU}{Division of Nuclear Physics, Department of Physics and 
 Astronomy, Uppsala University, Box 516, 75120 Uppsala, Sweden}
\newcommand*{\ASWarsN}{Nuclear Physics Division, National Centre for  Nuclear Research, ul.\ Pasteura 7, 02-093, Warsaw, Poland}
\newcommand*{\IPJ}{M. Smoluchowski Institute of Physics, Jagiellonian University, prof.\ 
 Stanis{\l}awa {\L}ojasiewicza~11, 30-348 Krak\'{o}w, Poland}
\newcommand*{\Edinb}{School of Physics and Astronomy, The University of
  Edinburgh, James Clerk Maxwell Building, Peter Guthrie Tait Road, Edinburgh
  EH9 3FD, Great Britain}
\newcommand*{\MS}{Institut f\"ur Kernphysik, Westf\"alische 
 Wilhelms--Universit\"at M\"unster, Wilhelm--Klemm--Str.~9, 48149 M\"unster, 
 Germany}
\newcommand*{\ASWarsH}{High Energy Physics Division, National Centre for  Nuclear Research, ul.\ Pasteura 7, 02-093, Warsaw, Poland}
\newcommand*{\Budker}{Budker Institute of Nuclear Physics of SB RAS, 
 11~Acad.\ Lavrentieva Pr., Novosibirsk, 630090 Russia}
\newcommand*{\PGI}{Peter Gr\"unberg Institut, PGI--6 Elektronische 
 Eigenschaften, Forschungszentrum J\"ulich, 52425 J\"ulich, Germany}
\newcommand*{\DUS}{Institut f\"ur Laser-- und Plasmaphysik, Heinrich Heine 
 Universit\"at D\"usseldorf, Universit\"atsstr.~1, 40225 Düsseldorf, Germany}
\newcommand*{\IFJ}{The Henryk Niewodnicza{\'n}ski Institute of Nuclear 
 Physics, Polish Academy of Sciences, ul.\ Radzikowskiego~152, 31-342 
 Krak\'{o}w, Poland}
\newcommand*{\PITue}{Physikalisches Institut, Eberhard Karls Universit\"at 
 T\"ubingen, Auf der Morgenstelle~14, 72076 T\"ubingen, Germany}
\newcommand*{\Kepler}{Kepler Center for Astro and Particle Physics,
  Physikalisches Institut der Universit\"at T\"ubingen, Auf der 
 Morgenstelle~14, 72076 T\"ubingen, Germany}
\newcommand*{\IKPJ}{Institut f\"ur Kernphysik, Forschungszentrum J\"ulich, 
 52425 J\"ulich, Germany}
\newcommand*{\ZELJ}{Zentralinstitut f\"ur Engineering, Elektronik und 
 Analytik, Forschungszentrum J\"ulich, 52425 J\"ulich, Germany}
\newcommand*{\Erl}{Physikalisches Institut, Friedrich--Alexander Universit\"at
 Erlangen--N\"urnberg, Erwin--Rommel-Str.~1, 91058 Erlangen, Germany}
\newcommand*{\ITEP}{National Research Centre "Kurchatov Institute", Moscow, 123182 Russia}
\newcommand*{\Giess}{II.\ Physikalisches Institut, 
 Justus--Liebig--Universit\"at Gie{\ss}en, Heinrich--Buff--Ring~16, 35392 
 Giessen, Germany}
\newcommand*{\IITI}{Discipline of Physics, Indian Institute of Technology 
 Indore, Khandwa Road, Indore, Madhya Pradesh 453 552, India}
\newcommand*{\HepGat}{High Energy Physics Division, Petersburg Nuclear Physics 
 Institute named by B.P.\ Konstantinov of National Research Centre ``Kurchatov 
 Institute'', 1~mkr.\ Orlova roshcha, Leningradskaya Oblast, Gatchina, 188300
 Russia}
\newcommand*{\HeJINR}{Veksler and Baldin Laboratory of High Energiy Physics, 
 Joint Institute for Nuclear Physics, 6~Joliot--Curie, Dubna, 141980 Russia}
\newcommand*{\Katow}{August Che{\l}kowski Institute of Physics, University of 
  Silesia, ul.\ 75 Pu{\l}ku Piechoty 1, 41-500 Chorz\'{o}w, Poland}
\newcommand*{\NITJ}{Department of Physics, Malaviya National Institute of 
 Technology Jaipur, JLN Marg, Jaipur, Rajasthan 302 017, India}
\newcommand*{\Bochum}{Institut f\"ur Experimentalphysik I, Ruhr--Universit\"at 
 Bochum, Universit\"atsstr.~150, 44780 Bochum, Germany}
\newcommand*{\IITB}{Department of Physics, Indian Institute of Technology 
 Bombay, Powai, Mumbai, Maharashtra 400 076, India}
\newcommand*{\Tomsk}{Department of Physics, Tomsk State University, 36~Lenin 
 Ave., Tomsk, 634050 Russia}
\newcommand*{\KEK}{High Energy Accelerator Research Organisation KEK, Tsukuba, 
 Ibaraki 305--0801, Japan} 
\newcommand*{\ASLodz}{Astrophysics Division, National Centre for Nuclear
 Research, Box~447, 90-950 {\L}\'{o}d\'{z}, Poland}
\newcommand*{\York}{School of Physics Engineering and Technology, The University of York, Heslington, York, YO10 5DD, UK}
\newcommand*{\mail}{barbara.klos@us.edu.pl}

\author{P.~Adlarson}    \affiliation{\IKPUU}
\author{W.~Augustyniak} \affiliation{\ASWarsN}
\author{W.~Bardan}      \affiliation{\IPJ}
\author{M.~Bashkanov}   \affiliation{\York}
\author{F.S.~Bergmann}  \affiliation{\MS}
\author{M.~Ber{\l}owski}\affiliation{\ASWarsH}
\author{A.~Bondar}      \affiliation{\Budker}
\author{M.~B\"uscher}   \affiliation{\PGI}\affiliation{\DUS}
\author{H.~Cal\'{e}n}   \affiliation{\IKPUU}
\author{I.~Ciepa{\l}}   \affiliation{\IFJ}
\author{H.~Clement}     \affiliation{\PITue}\affiliation{\Kepler}
\author{E.~Czerwi{\'n}ski}\affiliation{\IPJ}
\author{K.~Demmich}     \affiliation{\MS}
\author{R.~Engels}      \affiliation{\IKPJ}
\author{A.~Erven}       \affiliation{\ZELJ}
\author{W.~Eyrich}      \affiliation{\Erl}
\author{P.~Fedorets}    \affiliation{\ITEP}
\author{K.~F\"ohl}      \affiliation{\Giess}
\author{K.~Fransson}    \affiliation{\IKPUU}
\author{F.~Goldenbaum}  \altaffiliation[present address: ]{\GSI}\affiliation{\IKPJ}
\author{A.~Goswami}     \affiliation{\IKPJ}\affiliation{\IITI}
\author{K.~Grigoryev}   \altaffiliation[present address: ]{\GSI}\affiliation{\IKPJ}
\author{L.~Heijkenskj\"old}\altaffiliation[present address: ]{\Mainz}\affiliation{\IKPUU}
\author{V.~Hejny}       \affiliation{\IKPJ}
\author{L.~Jarczyk}     \altaffiliation{deceased}\affiliation{\IPJ}
\author{T.~Johansson}   \affiliation{\IKPUU}
\author{B.~Kamys}       \affiliation{\IPJ}
\author{G.~Kemmerling}\altaffiliation[present address: ]{\JCNS}\affiliation{\ZELJ}
\author{A.~Khoukaz}     \affiliation{\MS}
\author{A.~Khreptak}    \affiliation{\IPJ}
\author{D.A.~Kirillov}  \affiliation{\HeJINR}
\author{S.~Kistryn}     \affiliation{\IPJ}
\author{H.~Kleines}\altaffiliation[present address: ]{\JCNS}\affiliation{\ZELJ}
\author{B.~K{\l}os}    \altaffiliation[Corresponding author: ]{\mail} \affiliation{\Katow}
\author{W.~Krzemie{\'n}}\affiliation{\ASWarsH}
\author{P.~Kulessa}     \affiliation{\IFJ}
\author{A.~Kup\'{s}\'{c}}\affiliation{\IKPUU}\affiliation{\ASWarsH}
\author{K.~Lalwani}     \affiliation{\NITJ}
\author{D.~Lersch}\altaffiliation[present address: ]{\FSU}\affiliation{\IKPJ}
\author{B.~Lorentz}    \altaffiliation[present address: ]{\GSI}\affiliation{\IKPJ}
\author{A.~Magiera}     \affiliation{\IPJ}
\author{R.~Maier}       \affiliation{\IKPJ}
\author{P.~Marciniewski}\affiliation{\IKPUU}
\author{B.~Maria{\'n}ski}\altaffiliation{deceased}\affiliation{\ASWarsN}
\author{H.--P.~Morsch}  \affiliation{\ASWarsN}
\author{P.~Moskal}      \affiliation{\IPJ}
\author{W.~Parol}       \affiliation{\IFJ}
\author{E.~Perez del Rio}\affiliation{\IPJ}
\author{N.M.~Piskunov}  \affiliation{\HeJINR}
\author{D.~Prasuhn}     \affiliation{\IKPJ}
\author{D.~Pszczel}     \affiliation{\IKPUU}\affiliation{\ASWarsH}
\author{K.~Pysz}        \affiliation{\IFJ}
\author{J.~Ritman} \altaffiliation[present address: ]{\GSI}\affiliation{\IKPJ}\affiliation{\Bochum}
\author{A.~Roy}         \affiliation{\IITI}
\author{O.~Rundel}      \affiliation{\IPJ}
\author{S.~Sawant}      \affiliation{\IITB}
\author{S.~Schadmand}   \altaffiliation[present address: ]{\GSI}\affiliation{\IKPJ}
\author{T.~Sefzick}  \altaffiliation[present address: ]{\GSI}\affiliation{\IKPJ}
\author{V.~Serdyuk}     \affiliation{\IKPJ}
\author{B.~Shwartz}     \affiliation{\Budker}
\author{T.~Skorodko}\affiliation{\PITue}\affiliation{\Kepler}
\author{M.~Skurzok}     \altaffiliation[present address: ]{\INFN}\affiliation{\IPJ}
\author{J.~Smyrski}     \affiliation{\IPJ}
\author{V.~Sopov}       \altaffiliation{deceased}\affiliation{\ITEP}
\author{R.~Stassen}     \affiliation{\IKPJ}
\author{J.~Stepaniak}   \affiliation{\ASWarsH}
\author{E.~Stephan}     \affiliation{\Katow}
\author{G.~Sterzenbach} \affiliation{\IKPJ}
\author{H.~Stockhorst}  \affiliation{\IKPJ}
\author{H.~Str\"oher}   \affiliation{\IKPJ}
\author{A.~Szczurek}    \affiliation{\IFJ}
\author{A.~Trzci{\'n}ski}\altaffiliation{deceased}\affiliation{\ASWarsN}
\author{M.~Wolke}       \affiliation{\IKPUU}
\author{A.~Wro{\'n}ska} \affiliation{\IPJ}
\author{P.~W\"ustner}   \affiliation{\ZELJ}
\author{A.~Yamamoto}    \affiliation{\KEK}
\author{J.~Zabierowski} \affiliation{\ASLodz}
\author{M.J.~Zieli{\'n}ski}\affiliation{\IPJ}
\author{J.~Z{\l}oma{\'n}czuk}\affiliation{\IKPUU}
\author{P.~{\.Z}upra{\'n}ski}\affiliation{\ASWarsN}
\author{M.~{\.Z}urek}   \affiliation{\ARG}

\newcommand*{\Mainz}{Institut f\"ur Kernphysik, Johannes 
 Gutenberg Universit\"at Mainz, Johann--Joachim--Becher Weg~45, 55128 Mainz, 
 Germany}
\newcommand*{\JCNS}{J\"ulich Centre for Neutron Science JCNS, 
 Forschungszentrum J\"ulich, 52428 J\"ulich, Germany}
\newcommand*{\FSU}{Department of Physics, Florida State University,
  77~Chieftan Way, Tallahassee, FL~32306-4350, USA}
\newcommand*{\INFN}{INFN, Laboratori Nazionali di Frascati, Via E. Fermi, 40, 
 00044 Frascati (Roma), Italy}
\newcommand*{\ARG}{Physics Division, Argonne National Laboratory, 9700 S Cass Ave, Lemont, IL 60439, USA}
\newcommand*{\GSI}{GSI Helmholtzzentrum für Schwerionenforschung GmbH, Planckstraße 1, 64291 Darmstadt, Germany}

\collaboration{WASA-at-COSY Collaboration}\altaffiliation{the author list reflects the once-existing composition of the WASA-at-COSY Collaboration,
within which the data presented in this paper have been obtained}

\author{A.~Deltuva}
\affiliation{Institute of Theoretical Physics and Astronomy, Vilnius University, Saulėtekio al. 3, LT-10257 Vilnius, Lithuania}
 \author{J.~Golak}\affiliation{\IPJ}
 
\author{A. Kozela}\affiliation{\IFJ}
\author{A. \L obejko}\affiliation{Institute of Experimental Physics, Faculty of Mathematics, Physics and Informatics, University of Gda\'nsk, PL-80308 Gda\'nsk, Poland}
 \author{P. U. Sauer}
\affiliation{Institute for Theoretical Physics, Leibniz University Hannover, D-30167 Hannover, Germany}
 \author{R.~Skibi\'nski}\affiliation{\IPJ}

\author{I.~Skwira-Chalot}
\affiliation{Faculty of Physics, University of Warsaw,PL-02093 Warszawa, Poland}
\author{A.~Wilczek}
\affiliation{\Katow}
\author{H.~Wita\l a} \affiliation{\IPJ}

\date{\today}

\begin{abstract}

The differential cross section for the $^{1}$H$(d,pp)n$ breakup reaction at deuteron beam energy 
of 380~MeV has been determined with high precision for  189 angular configurations of outgoing protons in the region of forward laboratory angles.  
The cross section data were compared to theoretical predictions based on the state-of-the-art
  nucleon-nucleon 
  potentials, combined with a three-nucleon force, the Coulomb interaction or carried out in a 
  relativistic approach. In the region of the lowest differential cross section, the discrepancy between the data and the theoretical predictions is observed, also in the case of relativistic calculations. 
\end{abstract}


\maketitle

\section{\label{secI}Introduction}

Properties of nuclei  at medium energies are determined to a large extent by
pairwise nucleon-nucleon (NN) interactions, which are a dominant component of the nuclear
 potential. However,  neglecting internal degrees of
 freedom of interacting nucleons, it is necessery to introduce the many-body forces,
 in particular the three-nucleon force (3NF).
  State-of-the-art models of 3NF's, like 
TM99 ~\cite{Coon2001},  Urbana~IX~\cite{Pudliner1997}, or Illinois \cite{Pieper2001},  
combined with the realistic  nucleon-nucleon (2N)  potentials, constitute the basis for  
calculations of binding energies and scattering observables.   Chiral  Effective Field Theory (ChEFT) 
provides a systematic 
construction of nuclear forces  in a fully consistent way: the 3N forces
 appear naturally at a certain order \cite{Epelbaum2009,Machleidt2016}. The  theoretical   calculations 
 including semi-phenomenological 3NF or 3NF stemming from 
 ChEFT,  reproduce with high accuracy binding energies of light 
 nuclei, see Refs.~\cite{Carlson2015, Piarulli2018, Maris2022} and references therein.  
 
 A comprehensive description of nuclear forces requires an accurate understanding of the few-
nucleon system dynamics. In particular, the three-nucleon system is the simplest one to test the 
 interaction models and constrain the 3NF parameters. 
 It was demonstrated that by including 3NF the description 
 of differential 
 cross section for elastic nucleon-deuteron  scattering is significantly improved as compared to 
 the calculations 
 based on NN interactions only \cite{Witala1998, Hatanaka2002, Mermod2004}. The remaining discrepancy between the 
 data and theoretical calculations appears above 100 MeV/nucleon and has not been explained by 
 either the influence of Coulomb interactions or relativistic effects, both of which are 
 predicted to be small and very local.
Studies of the $^{1}$H$(d,pp)n$ and $^{2}$H$(p,pp)n$ breakup reactions make important 
contribution to investigations of the 3NF. The advantage relies on the kinematic 
richness of the three-body  final state. There are experimental evidences of 
significant 3NF contributions 
to the differential cross section for the breakup reaction, starting already at relatively 
low beam energy of 65 MeV/nucleon \cite{Kistryn2005, Kistryn2013}. In contrast to the elastic scattering case, 
Coulomb interaction  is a very important component of the breakup reaction dynamics. 
The Coulomb interaction between protons modifies  the cross section data  over a 
 significant part of the phase space, in particular at forward laboratory 
angles of the $^{1}$H$(d,pp)n$ reaction \cite{Kistryn2006, Ciepal2015, Skwira_2024}. 
   At present, the Coulomb interaction and 3N forces are both included 
  into theoretical calculations and their interplay  can be studied \cite{Deltuva2005, Deltuva2006, Deltuva2009}. 

The relativistic calculations of the differential cross section for  breakup reaction 
led to different results  than the non-relativistic ones \cite{Witala2011}.  Due to the predicted significant
3NF effects and relativistic effects in the energy region between 150 and 200 MeV/nucleon \cite{Skibinski2006, Witala2011}, 
the question arises about their interplay. So far, no calculations have been performed
that treat in fully relativistic way both the NN and 3NF interactions. Under such circumstances one has to
rely on  systematic (in beam energy) studies over large phase space regions, with the 
aim to single out both   contributions on the basis of their different 
kinematic dependencies. 

 The very limited experimental data set in this range, especially when it comes to data covering 
 wide ranges of phase space, became the motivation for the experiments described here. 
  The  exception, a measurement of the $^{2}$H$(p,pp)n$ at 190\,MeV~\cite{Mardanpour_PhD, Mardanpour08, Hajar2021}, provided hints for  
  deficiencies in the description of  the cross section for the deuteron breakup 
reaction, even when 3NF is included.  The problem can be interpreted either as 
confirmation of the above mentioned issues in elastic scattering, 
or as a consequence of relativistic effects. Data presented in this work were taken in the 
inverse kinematics,  $^{1}$H$(d,pp)n$, complementing the studied region of the reaction 
phase space. They were part of a systematic study conducted with the WASA detector at COSY beam  in the energy range from 150 
to 200 MeV/nucleon, from which data at 170 MeV/nucleon have been published so far \cite{WASA2020}.

\section{\label{secII}Experiment and data analysis}
  
An experiment to investigate the $^{1}$H$(d,pp)n$ breakup reaction  using a~deu\-teron beam of 
150, 170, 190, 200\,MeV/nucleon was  performed in January 2013 at the 
Cooler Synchrotron COSY-J\"{u}lich. Energies 170, 190 and 200\,MeV/nucleon of the
 deuteron beam were changed in a supercycle mode of time length 30\,s, while 
 measurement at 150\,MeV/nucleon  was performed separately. Due to close to $4\pi$ acceptance and 
 moderate detection  threshold of the  WASA (Wide Angle Shower Apparatus) detection system,  
 differential cross  section data have been collected in a large part of  the breakup reaction  
 phase space. In this paper, the results of an analysis of the data collected at the beam 
energy of 190\,MeV/nucleon are shown, with focus on the proton-proton coincidences
 registered in the Forward Detector.    

\subsection{\label{secIIA} WASA detector}

\begin{figure}[h]
\centering
\includegraphics[width=85mm]{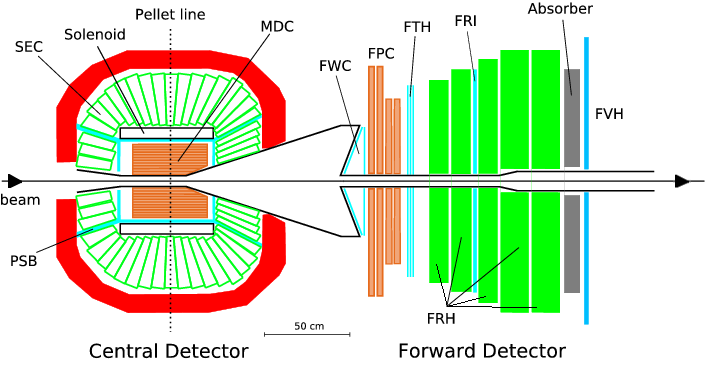}
\caption{(Color online) Schematic view of the detection system. }
\label{det}      
\end{figure}

The WASA  detector \cite{Bargholtz2008, WASA2004} (see Fig.~\ref{det}) consisted of 
 four main components: Central Detector (CD), Forward Detector (FD), Pellet Target and  
 Scattering Chamber. The pellet target  provided a narrow stream of frozen 
 hydrogen droplets. The Central Detector operated in the experiment described here, 
 but the present data analysis is limited to the registration of protons and deuterons only in the Forward Detector.

FD 
consisted of a 
set of detectors for the identification of charged hadrons and track reconstruction:
Forward Window Counter (FWC), Forward Proportional Chamber (FPC) and plastic scintillators: 
Forward Trigger  Hodoscope (FTH), Forward Range Hodoscope (FRH) and Forward Veto Hodoscope (FVH). 
 FPC was used for precise determination of particle emission angles. 
 The FD plastic scintillators were used for particle 
 identification and particle energy measurement. They all provided information for the first-level trigger logic; for details see Ref.~\cite{WASA2020}.
 Between the second (FRH2) and third (FRH3) layers of FRH there were two layers of Forward Range
Interleaving  Hodoscope (FRI). In addition, between the layers FRH3 and FRH4, a passive 
aluminum layer was placed to extend the energy range of particles stopped in the WASA detector not shown in Fig.~\ref{det}). 
  
\subsection{\label{secIIB}Data analysis}

The data analysis presented in this work was focused on proton-proton coincidences from the 
$^{1}$H$(d,pp)n$  breakup reaction at 190\,MeV/nucleon, registered  in the Forward Detector. This was a follow-up of the analysis performed for data collected at 
 170~MeV/nucleon, described in detail in Ref.~\cite{WASA2020}, thus  
 the data analysis procedures were similar. Below we will present the main steps and  focus 
on describing the differences associated with the different range of energies of particles 
in the exit channel. The multilayer structure of the Forward detector had a significant impact on the analysis 
procedures:  as the energy of the outgoing particles increased, more layers were involved in the 
track reconstruction, which affected both energy calibration and detection efficiency. 

\begin{figure}[h]
\begin{center}
\subfigure{
\includegraphics[width=8cm]{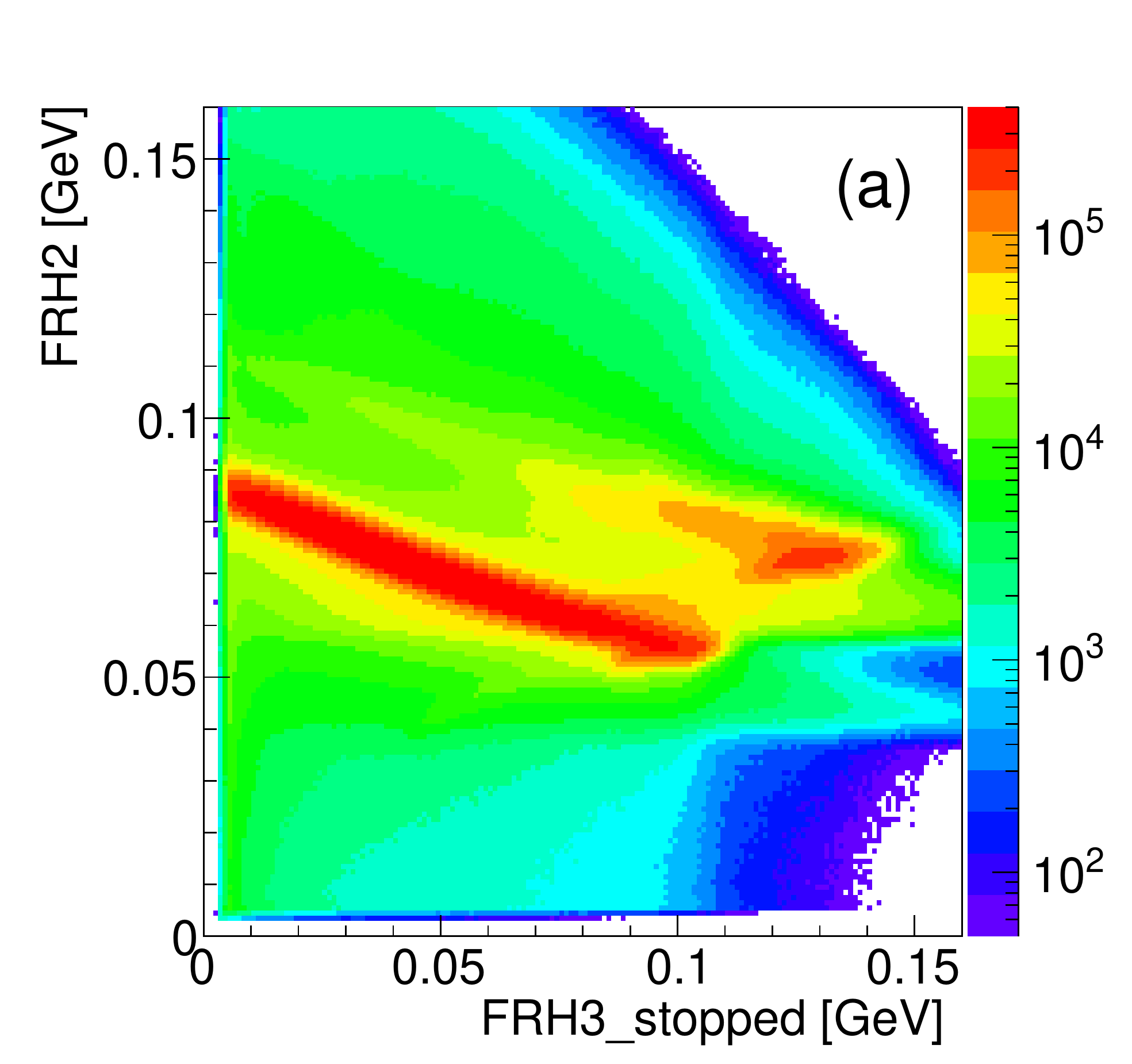}
}
\subfigure{
\includegraphics[width=8cm]{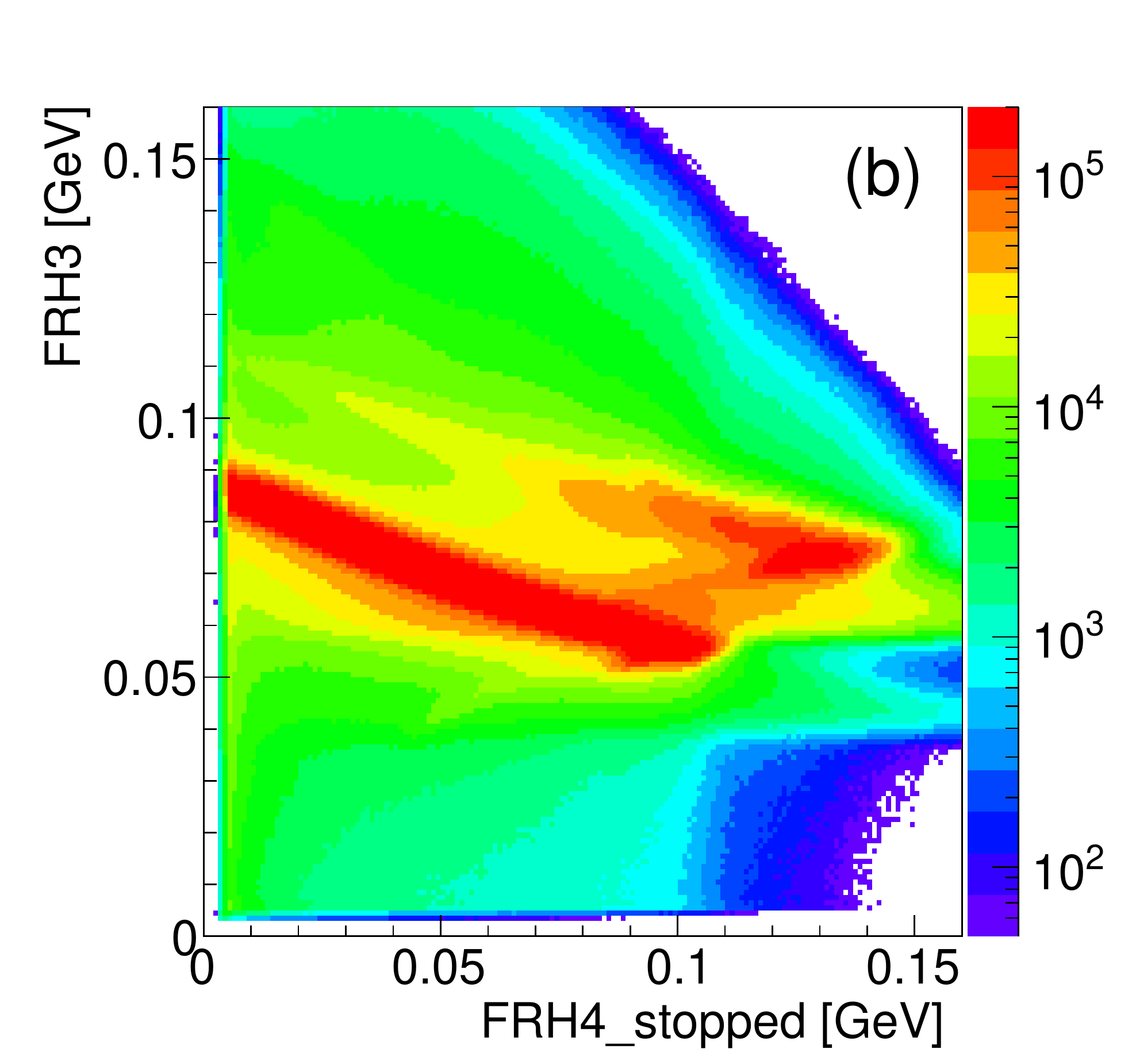}
}\\
\end{center}

\vspace{-0.7cm}

\caption{Particle identification spectra of particles stopped in  (a) 3rd layer and (b) 4th layer of FRH. Loci of protons and deuterons are clearly visible. }
\label{fig-pid}      
\end{figure}

In the first step of analysis, events 
of interest were selected: two protons from the breakup process and deuterons from elastic 
scattering channel, the latter used for the cross section normalization. The particle 
identification was based on the $\Delta E$-$E_R$ technique, where $E_R$ was the remaining energy deposit in the layer where the particle was stopped and $\Delta E$ denoted the energy deposit in the preceding layer. In the entire range of energies,
a clear separation between loci of protons and deuterons was observed, see examples in Fig.~\ref{fig-pid}.

\subsubsection{Analysis of breakup events}

\begin{figure}[ht]
\begin{center}
\includegraphics[width=8cm]{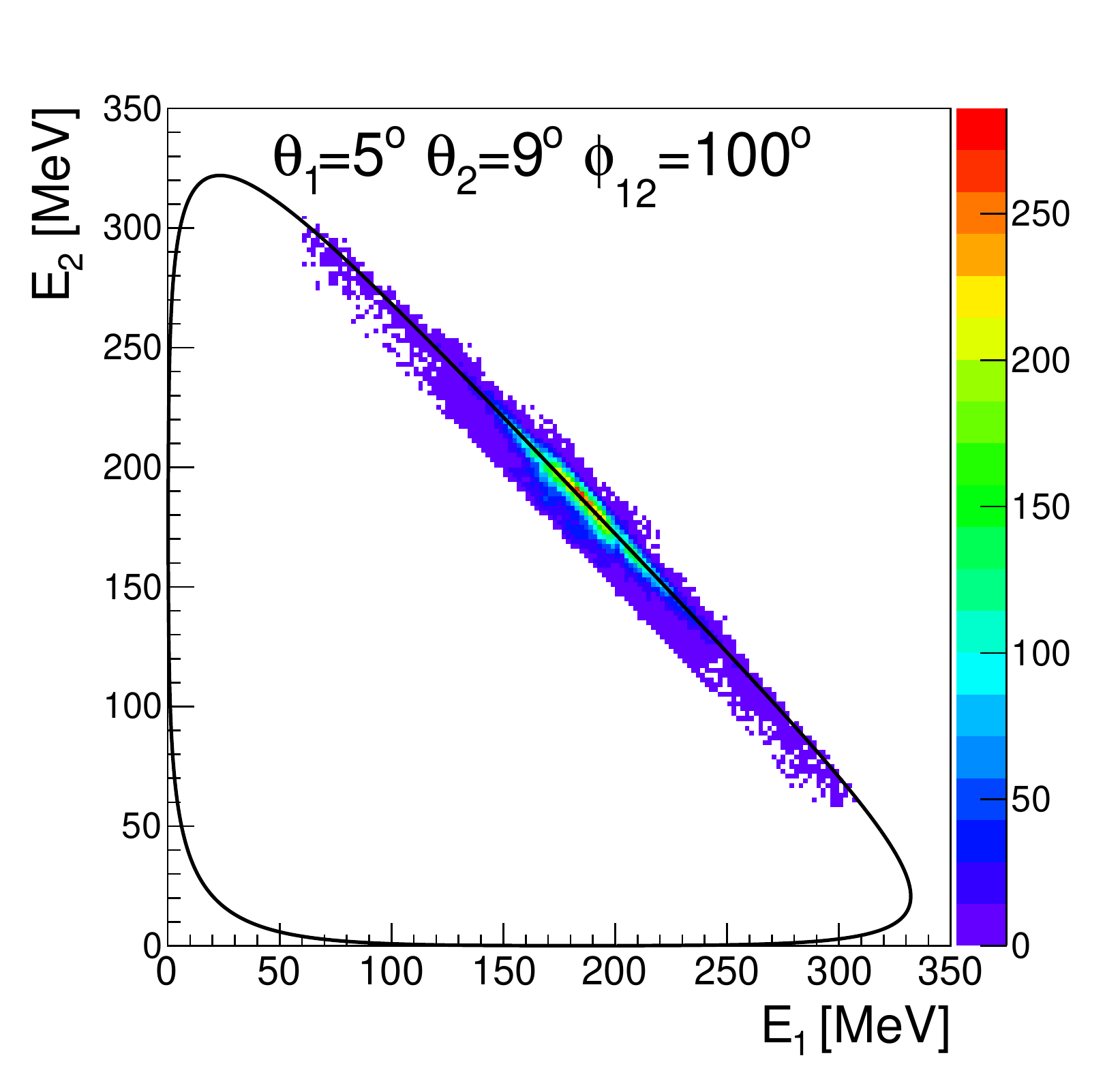}
\end{center}

\vspace{-0.5cm}

\caption{(Color online)  $E_1$ vs. $E_2$ coincidence spectrum of the two protons registered at $\theta _1$=5$^{\circ}\pm$ 1$^{\circ}$, $\theta _2$=9$^{\circ}\pm$ 1$^{\circ}$, and $\varphi _{12}$=100$^{\circ}\pm$ 5$^{\circ}$. The solid line shows a three-body kinematical curve, calculated for the central values of experimental angular ranges.  }
\label{kin}      
\end{figure}

The breakup configuration was defined by the emission angles of the two outgoing protons: two 
 polar angles $\theta _1$ and $\theta _2$ (with $\theta _1 \leq \theta_2$), between 5 and 15$^{\circ}$, and their relative azimuthal angle $\varphi _{12}$. 
 The data were integrated over the angular ranges of  $\theta _{1,2}$ ($\pm $1$^{\circ}$) and 
 $\varphi _{12}$ ($\pm$ 5$^{\circ}$). In order to reduce the background, cut on neutron peak in the missing mass spectrum \cite{Klos2022} was applied.  For each angular  configuration, the data were placed in~a kinematic spectrum showing the energy correlation of the proton pair, $E_1$ vs $E_2$ (see Fig.~\ref{kin}), and then transformed  
to the variable $S$, which shows the position along the kinematical curve~\cite{WASA2020}. The starting point of the $S$-curve ($S=0$) is chosen as a point for which energy of the second proton is minimal. This step required precise energy calibration, which was based on measurements of $dp$ elastic scattering at 
energies corresponding to minimum ionization~\cite{Vlasov_PhD}. Since the FRI detector was not used in a number of  
previous runs, its calibration was not included in the main calibration procedure and was 
known with lower accuracy. The appropriate corrections have been applied on the basis of energy deposited in the previous layer  but the energy 
resolution was diminished. The affected energy region corresponded 
to the slope of the cross section $S$-distribution~\cite{Klos2022} 
and the solution was to apply adequately wide binning in $S$ of 24~MeV.
A similar 
issue arose for particles 
 stopped in the aluminum layer between FRH3 
 and FRH4, and its impact on the reconstructed energy  was even more pronounced due to the considerable thickness of this passive layer.
 Since it affected only the tails of the $S$-distributions, it was decided to discard these regions, which 
 only slightly limited an $S$ range of the cross section data. 

The energy calibration was validated by the corelation of proton energies in each of the studied breakup configurations (see example in Fig.~\ref{kin}) and by reproduction of the neutron missing mass, for details see Ref.~\cite{Klos2022}.

WASA Monte Carlo program based on the Geant 3 simulation package was used for precise determination of efficiency of the detection 
system.
The analysis of generated data included the detector acceptance and all cuts 
applied in the analysis. The efficiency for registering and identifying elastically scattered 
deuterons was determined for each polar angle and typically reached  
about 80\%. The efficiency  for proton-proton coincidences was determined for each 
kinematical configuration with defined integration limits: 
$\Delta\theta_1$=$\Delta\theta_2$=2$^{\circ}$ and 
$\Delta\varphi _{12}$=10$^{\circ}$. Differently from the analysis performed at lower beam energy 
\cite{WASA2020}, a common efficiency factor did not apply 
for the entire  $S$-range. This was due to the difference in the number of detector layers penetrated by coincident protons which depended on their energy. In the central part of the $S$ distribution, typical efficiency for registering breakup events was  60\%, while on the tails, it was reduced down to 50\% due to one proton penetrating to the FRH3 layer~\cite{Klos2022}.

\subsubsection{Luminosity \label{normaliz}}

The absolute normalization was derived from the luminosity determined on 
the basis of the number of elastically scattered deuterons. Deuterons were sorted with respect to their laboratory scattering angles,  between 8$^{\circ}$ and 14$^{\circ}$ in a step of 1$^{\circ}$. To remove the tail associated with hadronic interactions, limits were set on the deuteron peak in the corresponding energy spectra.
After correcting for efficiency,  the resulting number of deuterons recorded at each angle was normalised based on the known elastic scattering cross section. 

\begin{figure}[h]
\begin{center}
\includegraphics[width=7.cm]{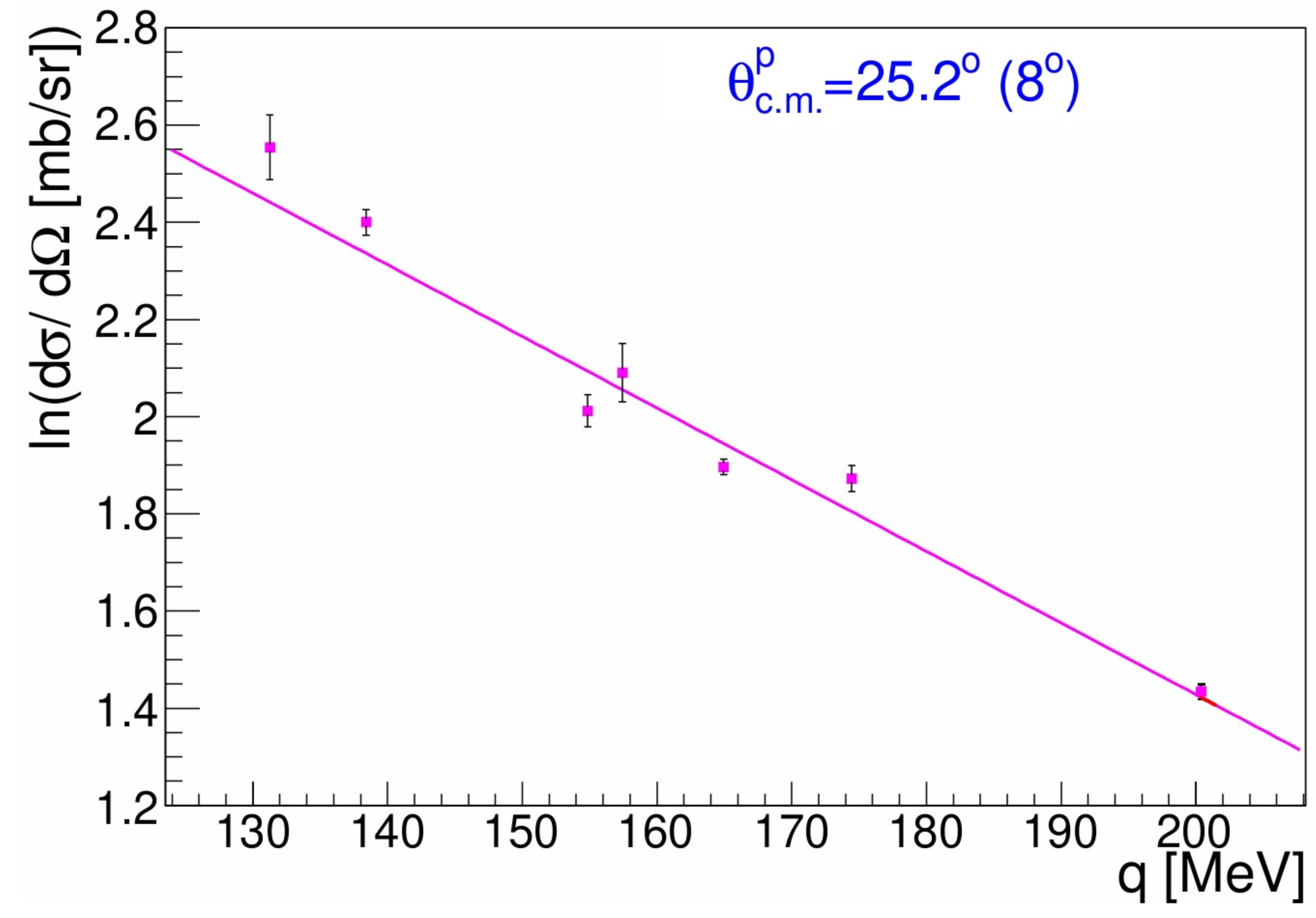}
\includegraphics[width=7.cm]{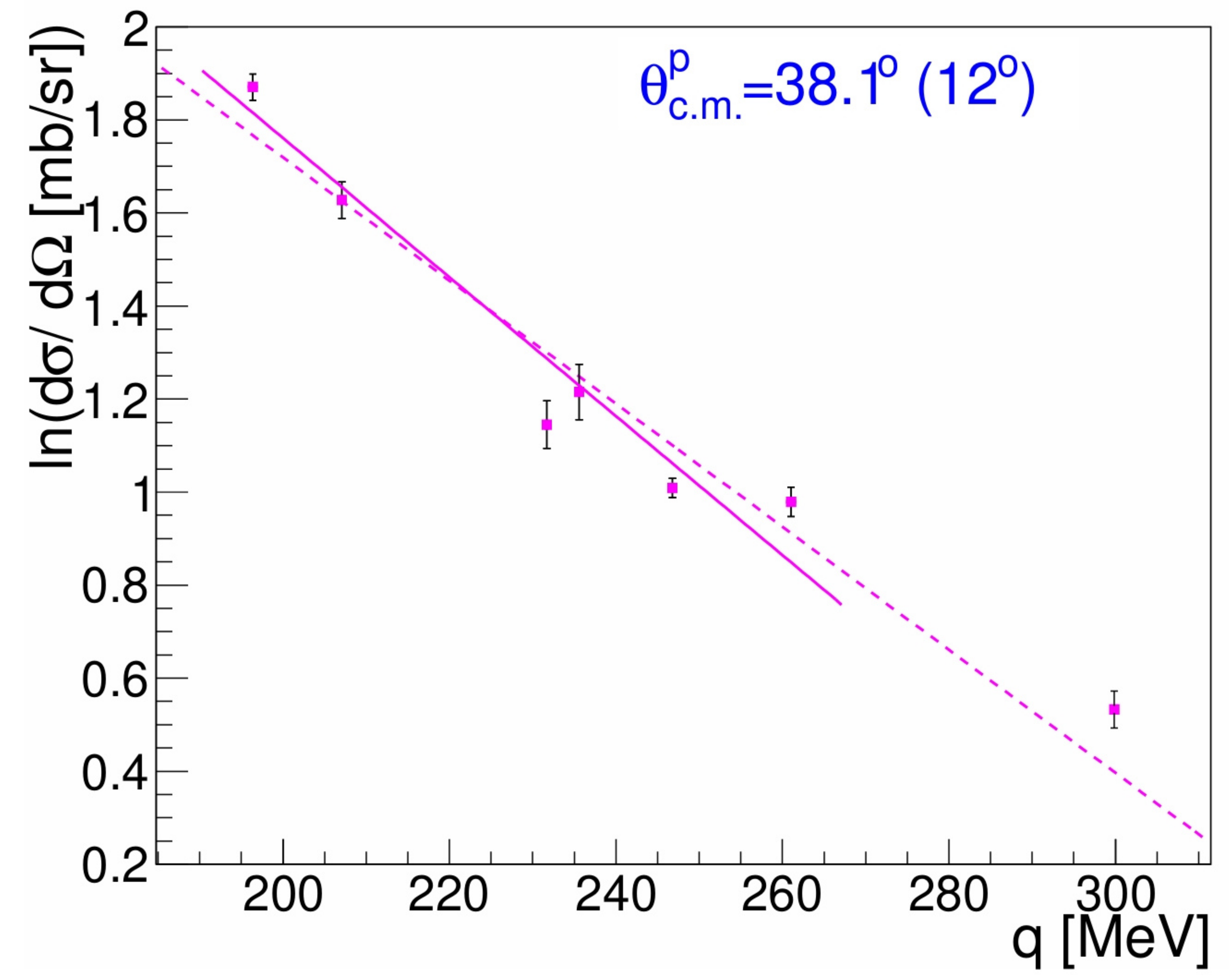}
\end{center}
\caption{(Color online)  Experimental differential cross section of the deuteron-proton elastic scattering  at the incident-beam energies: 108\,MeV, 120\,MeV, 150\,MeV, 170\,MeV, 190\,MeV \cite{Ermisch2003,Ermisch2005}, 155\,MeV \cite{Kuroda64} and 250 MeV \cite{Hatanaka2002} presented as ln($\sigma$) in a function of $q$ for two scattering angles, specified in the panels. Two lines in the lower panel show linear fits including (dashed line) or excluding (solid line) the point for the highest beam energy of 250\,MeV. } 
\label{q_fit}      
\end{figure}

Due to the considerable scatter of existing experimental data in the intermediate energies, a previously used strategy of a fit to a wide range of beam energies \cite{WASA2020} was pursued. In the current analysis we also used the fact, that  the angular range of elastically scattered deuterons  corresponds to  $\theta^{p}_{CM}<$50$^{\circ}$, which  is a region of a so-called soft-scattering.
In the elastic scattering, a part of the momentum and energy is transferred from the beam to the target, thus the square of the four-momentum transfer $t<0$,  and it can be described by: 
\begin{equation}
t = -2 {\rm p^{*2}}c^2(1-\cos\theta_{CM}) ,
 \label{telastic}
\end{equation}
with  ${\rm p^*}$ corresponding to the momentum and $\theta_{CM}$ denoting the beam particle scattering angle in the center of mass system. A four-momentum transfer defined as $qc = \sqrt{-t}$  is related to the momentum transferred to the target and reflects the “hardness” of the collision. The soft collisions mentioned above are long-range
collisions characterized by small $q$. This region is also called the  \emph{scaling 
region} and it is dominated by the direct term in the two-nucleon potential. Moreover, the 
dynamical part (after removing the phase-space factor) of
the cross section predominately depends on the momentum transfer and has a very
small dependence on the beam energy (for further discussion see Ref.~\cite{Ahmed_PhD}). The cross section dependence on the momentum transfer has dominating exponential character. For beam energies ranging between 108 and 250 MeV  and   momentum transfers $q<280$~MeV/c, the 
 $\ln \sigma_{CDB+TM99} $ calculated with the CD-Bonn potential combined with the TM99 3NF reveal a linear dependence on $q$.  At $ \theta_{CM}>45^{\circ} $ and  $q>300$ MeV/c small  departure from linearity  can be observed. This validates the linear fit to the
experimental data, with the exception of the highest $q$'s. 
In Fig.~\ref{q_fit}, the available $pd$ elastic cross data and the corresponding fits are shown for two deuteron scattering angles.  

\begin{figure}[h]
\begin{center}
\includegraphics[width=8.5cm]{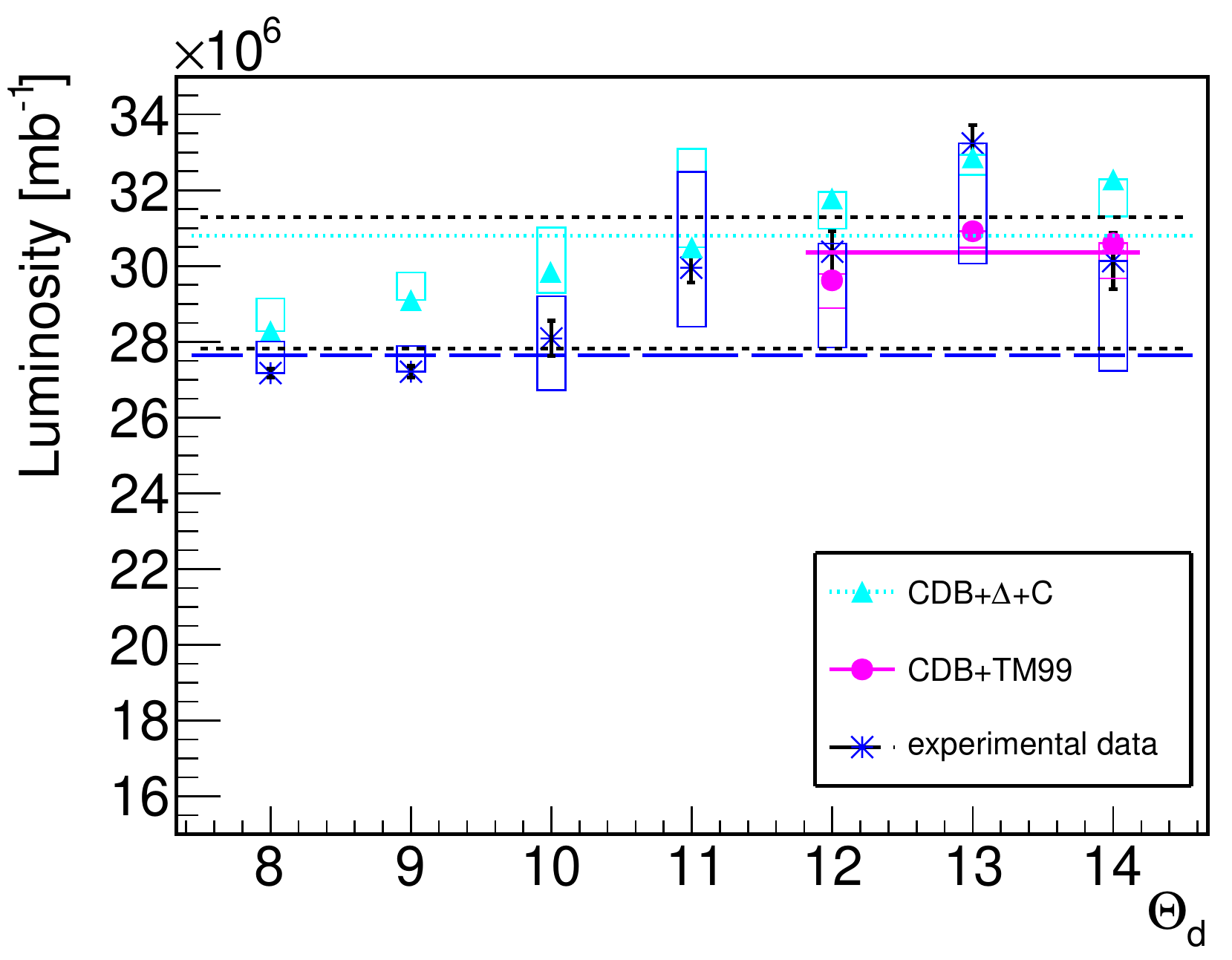}
\end{center}
\vspace{-0.6cm}
\caption{(Color online) Determination of the integrated luminosity. The integrated luminosity values were calculated  independently for each deuteron scattering angle.  Each set of points and corresponding average (horizontal line) corresponds to a different way of determining  elastic scattering cross section used for normalization, as specified in the legend. Two additional dashed lines present the systematic uncertainty of the final result. For more detais see the text.}
\label{fig-luminosity}      
\end{figure}

 Similarly to the previous analysis~\cite{WASA2020}, the fit results were were cross-checked with theoretical calculations. In the soft scattering region, the theoretical calculations including 3NF, for example CDB+$\Delta$ or CDB+TM99 (see Sec.~\ref{theory}) provide adequate description of the data. At the lowest $q$ one should however take into account Coulomb interactions modifying the cross section. Therefore, two types of calculations were taken into account: the one including Coulomb, CDB+$\Delta$+C, applicable in the full range of studied scattering angles, and CDB+TM99, to be used in the range of laboratory angles between 12$^{\circ}$ and 14$^{\circ}$ where the predicted Coulomb effects are below 2\% (as stems from the comparison of the CDB+$\Delta$+C and CDB+$\Delta$ predictions). The calculated cross section values were used for determining integrated luminosity. The corresponding data points (cyan triangles and magenta dots in Fig.~\ref{fig-luminosity}) have uncertainties resulting from uncertainties of the detection efficiency corrections. These results are compared to the luminosity values (blue stars) obtained with the experimental cross sections  resulting from the fits described above. In the latter case, uncertainties of the $\ln{\sigma}$ vs.~$q$ fits were included in the final uncertainty boxes. The spread of results within each of the methods illustrates differences of shapes of angular cross section distributions obtained in this experiment and the corresponding normalization cross sections. 

This spread sometimes exceeds the estimated systematic uncertainties of individual points. For each normalization method, the corresponding average was calculated, with weights that include systematic uncertainties (horizontal lines, as specified in the legend of Fig.~\ref{fig-luminosity}). 
Analogous to the previous analysis~\cite{WASA2020}, the result obtained with  CDB+TM99 potential is taken as a luminosity value (horizontal solid magenta line in Fig.~\ref{fig-luminosity}), and the other ones 
are used to determine systematic uncertainty. The systematic uncertainties marked as horizontal black dashed lines cover the range of scatter of individual data points obtained with all three normalization methods. Finally, the average integrated luminosity obtained for the full  set of data is (3.036 $\pm$ 0.003)~$\cdot$10$^{7}$mb$^{-1}$ with asymmetric systematic uncertainty of  (-0.254, +0.093)~$\cdot$10$^{7}$mb$^{-1}$, corresponding to a relative error of -8\% and +3\%.

\subsubsection{Experimental uncertainties } \label{error}

Statistical uncertainties of the measured cross-section values comprise an uncertainty of the measured 
number of the breakup coincidences and of the luminosity. In all 189 kinematic configurations 
statistical errors in peaks of the cross-section distributions are below 1\%. 

The data consistency checks and discussion of systematic uncertainties presented in 
Ref.~\cite{WASA2020} are valid also for the data set presented here.
The systematic uncertainties dominate over statistical ones and stem primarily from three sources: 
\begin{itemize}
     \item Luminosity determination, which affects the common normalization factor for all data points. The systematic uncertainty of the luminosity, as discussed in Sec.~\ref{normaliz}, is asymmetric and leads to uncertainty of the cross section: -3\% and +8\%. 
     \item Uncertainty of efficiency for detecting proton-proton coincidences. It corresponds to the accuracy of the MC simulations and consists of two parts: the systematic error of hadronic interaction modeling (7\%) and the statistical accuracy of the simulations; the latter one varies typically between  1\% and 7\% (only in 6 out of 189 configurations it exceeds 7 \%).  
    \item Background subtraction procedure -  the corresponding uncertainty was determined for each data point in two ways: varying the range around the gaussian peak where the linear background was determined, and checking gaussian shape of the distribution obtained after the background subtraction (by comparing the integral of the fitted Gauss with the sum of events);  in the central parts of the $S$-distributions it varies between 1 and 9\%.  
\end{itemize}

Finally, summing all the components in squares,  we obtain systematic uncertainties ranging between 8 and 16\%  for 
the central regions of the studied $S$ distributions, thus dominating  over the statistical ones. 
 
\section{Theoretical calculations} \label{theory}

The experimental cross section data measured at 190~MeV/nucleon have been compared to results of 
the same set of the calculations as in case of the earlier published data 
collected at 170 MeV/nucleon.  The list of theoretical 
approaches is given in Table~\ref{theo}, for details see Ref.~\cite{WASA2020} and references therein. To study the impact of 
individual contributions to the system dynamics, a comparison was first made with calculations 
based on pure NN potentials (2N and CDB in Table~\ref{theo}). Next, the 3NF was 
taken into account (2N+TM99, CDB+$\Delta$). The Coulomb interaction  was applied to a purely 
nucleonic CDB potential (CDB+C) and its coupled-channel extension (CDB+$\Delta$+C).
In one case, CDBrel, the relativistic effects were taken into account.
The group of calculations, CDB, CDB+TM99, 
2N, 2N+TM99, CDBrel, was performed by Wita\l a et al.,  
while the second group, CDB, CDB+$\Delta$ and CDB+$\Delta$+C, was performed by Deltuva.  

Prior to comparing with the data, the theoretical predictions were averaged 
 over the same angular bins as used in the data analysis ($\Delta\theta_{1,2}=2^{\circ}$, 
$\Delta\varphi_{12}=10^{\circ}$) and projected onto relativistic kinematics, see 
Ref.~\cite{Kistryn2013}. The relativistic calculations are the only exception: the calculations
were performed for central values of the angular ranges alone. 

\begin{table}[h]
\caption{Definition of abbreviations applied for naming theoretical calculations. ``aver'' means averaging over angular ranges accepted in data analysis.}\label{theo}
\vspace{-2mm}
\begin{small}
\begin{center}
\begin{tabular}{|l|l|c|l|}
\hline
abbreviation  & description & aver & Ref. \\
\hline
\hline
&  &  & \\
CDB &  CD Bonn potential & Yes & \cite{Machleidt1996,Machleidt2001} \\
 &   & & \\
 \hline
&  &  & \\
 &  CD Bonn potential &  & \cite{Coon1981,Coon2001,Glockle1981}\\
CDB+TM99 & combined with TM99 3NF  & Yes & \\
&   & & \\
\hline
 &  potentials:  &  & \\
2N &  AV18, CD Bonn,  & Yes & \cite{Machleidt1996,Machleidt2001,Wiringa1995,Stoks1994}\\
& Nijmegen I and II & & \\
\hline
 &  potentials  (as above) &  & \\
2N+TM99 & combined with TM99 3NF  & Yes & \cite{Coon1981,Coon2001,Glockle1981}\\
&   & & \\

\hline
& coupled-channel potential&  & \\
CDB+$\Delta$ & CD Bonn+explicit $\Delta$  & Yes & \cite{Deltuva2003a} \\
&  &  & \\
\hline
& coupled-channel potential &  & \\
CDB+$\Delta$+C &  CDB+$\Delta$   & Yes & \cite{Deltuva2005,Deltuva2006}\\
 & with Coulomb force   & & \\
\hline
&CD Bonn potential &  & \\
CDBrel & relativistic  & No & \cite{Skibinski2006}\\
 &  calculations  &  &\\
\hline
\end{tabular}
\end{center}
\end{small}
\end{table}

\indent
The most extensive developments of nuclear potentials are nowadays carried out within the 
framework of ChEFT.  
It has been shown, that with regard to NN interactions it is necessary to perform calculations at 5th 
 order ($\nu$=5), {\it i.e.} N4LO (see the discussion in \cite{Machleidt2016}). Thus, the 3NF at the same order is required for the consistency. So far the complete calculations for Nd system at N3LO are unavailable and only approaches based on
   the realistic potentials are considered in this work. Recent developments in calculating observables for the breakup reaction within ChEFT approach are discussed in Ref.~\cite{Maris2022}. 
 
\section{\label{secIV}Results}

\begin{center}
\begin{figure*}[th]
\includegraphics[width=15.5cm]{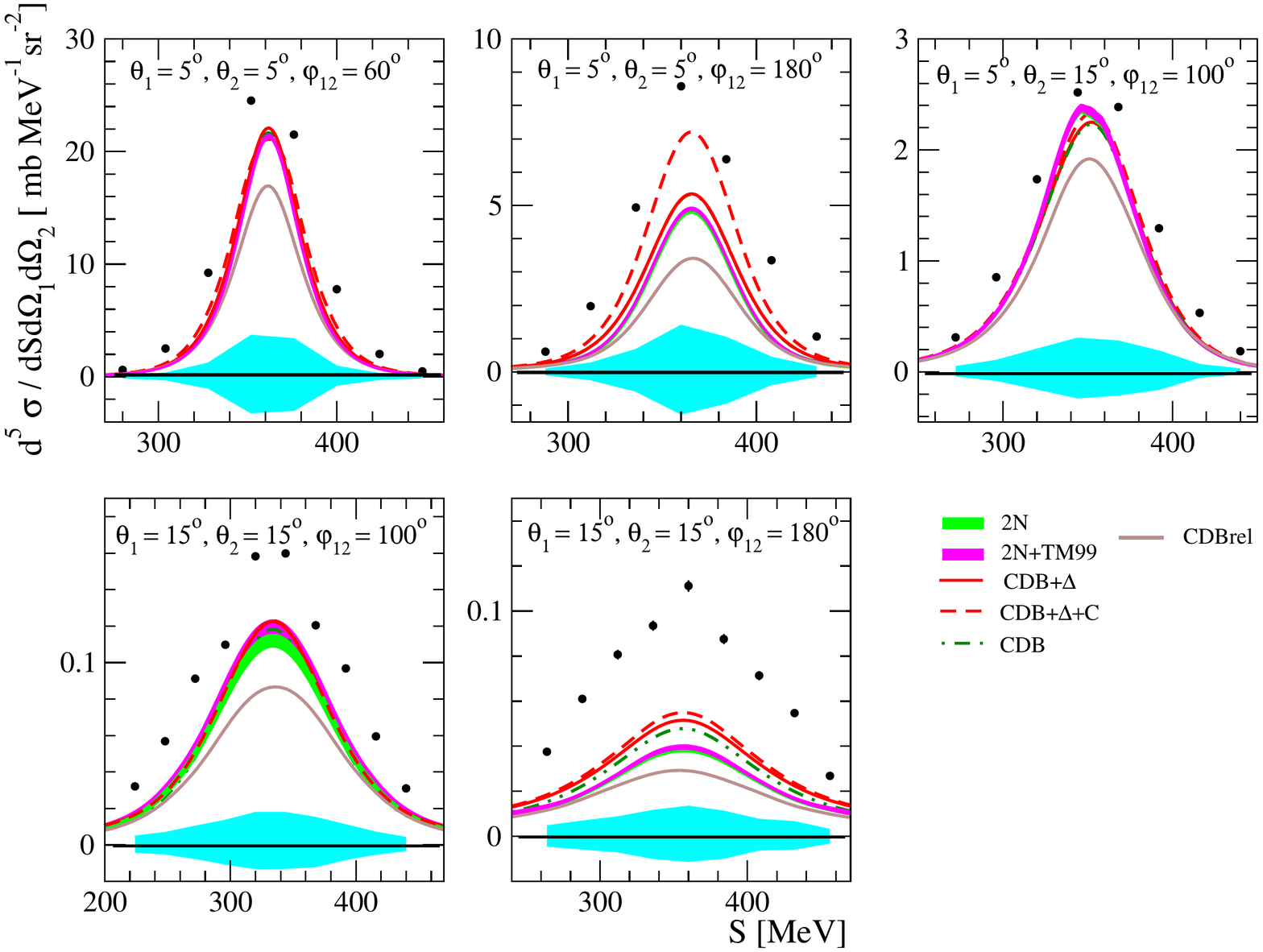}
\caption{(Color online) Differential cross section of $^{1}$H$(d,pp)n$ breakup reaction at beam energy of 190\,MeV/nucleon shown in a function of the $S$ variable for the set of configurations specified in the panels. Bars representing statistical uncertainties are usually smaller than data points. Systematic uncertainties are represented by bands. Data are compared to theoretical calculations specified in the legend. 
}
\label{kon2rel}      
\end{figure*}
\end{center}

The differential cross section for a regular grid of polar and azimuthal angles with a constant 
step in arclength variable $S$ (24 MeV) was experimentally obtained. Polar angles of the two protons, $\theta _1$ 
and $\theta _2$, were changed between 5$^\circ$ and 15$^\circ$ with the step size of 2$^\circ$ and 
their relative azimuthal angle  $\varphi _{12}$ was analyzed in the~range from 20$^\circ$ to 
180$^\circ$, with the step size of 20$^\circ$. In total, 189 configurations have been analysed. 
Fig.~\ref{kon2rel} presents cross section distributions for the  configurations for which relativistic  calculations are available.

 This set can be considered representative for the studied phase space,  with the smallest ($5^{\circ}$) and largest ($15^{\circ}$) among the analyzed polar angles, both symmetric and maximally asymmetric cases (respectively $\theta_2-\theta_1 = 0^\circ$ and $10^\circ$), combined with two very different relative azimuthal angles ($\varphi_{12}=60^\circ$ and $180^\circ$).
   The data are compared to all the theoretical calculations listed in Table~\ref{theo}. Theoretical bands correspond to non-relativistic 2N and 2N+TM99 calculations by Wita\l a et al.~and their width is related to differences between predictions obtained with individual NN potentials. Since the bands are narrow, further in  quantitative analyses only CDB and CDB+TM99 will be  used. In all the cases shown in  Fig.~\ref{kon2rel}, relativistic calculations  are significantly below the data.  The predicted 3NF effects are small or negligible.  In the top panels, a certain  underestimation of the measured cross section by theories is observed, but roughly contained within systematic uncertainties. This statement requires quantitative verification. In one case the Coulomb effect is clearly present.  
The configurations shown in the bottom panels, with $\theta_1=\theta_2 = 15^\circ$, are characterized by a very low cross section, one or two orders lower (in the peaks) than maximal cross section values  
in the top panels. In these cases  all the  calculations significantly 
underestimate the experimental data. This confirms the findings from the measurement at 170 MeV/nucleon \cite{WASA2020}, though for the discussed currently data at 190 MeV/nucleon, discrepancies are even more pronounced, reaching a factor of 2 in the best case, or even 4 in the case of the relativistic approach.

Figs.~\ref{fig5t5}~-~\ref{fig15t15} in Appendix~A present  the  differential cross section results for the full set of studied kinematic configurations.
A brief review of them leads to similar qualitative conclusions:
\begin{itemize}
\item 3NF effects are small in the entire angular region under study.
\item Coulomb effects, as expected, dominate in the most forward  configurations ($\theta_1,\theta_2= 5^\circ$ or $7 ^\circ$)  with the lowest ($\varphi_{12} = 20^\circ$) and highest ($\varphi_{12} \geq 140^\circ$) relative azimuthal angles. In these regions the calculations including Coulomb interaction provide more adequate results than other ones,  though in some cases the Coulomb effect is less pronounced in the data than predicted by theory.
\item The overall description of the data for configurations with $\varphi_{12} \leq 100^\circ$ is good, and while the experimental points lie  systematically above the theoretical lines, the differences are comparable to systematic uncertainties and may be due to  uncertainty of the luminosity. 
\item In configurations with the largest studied polar angles ($\theta_1=\theta_2 =13^\circ, 15 ^\circ$), nearly coplanar   ($\phi_{12} = 160^\circ, 180^\circ$), the theoretical predictions deviate very strongly from the data, underestimating them. 
 \end{itemize}
These observations will be further quantitatively discussed by means of a $\chi^2$ analysis.

\subsection{$\chi^2$ analysis}

Quantitative analysis of the description of the cross section data ($\sigma_{exp}$) 
provided by various calculations ($\sigma_{teor}$) is performed in 
terms of $\chi^2$-like variables. Due to the dominating contribution of systematic uncertainties, the following definition has been applied, which also facilitates direct comparisons with the results from Ref.~\cite{WASA2020}:

\begin{equation}
\chi^2 = \frac{1}{n_{d.o.f.}} \sum \frac{(\sigma_{\rm teor}(\xi)-\sigma_{\rm exp}(\xi))^2}{(\Delta\sigma_{\rm st}(\xi)+\Delta\sigma_{\rm sys}(\xi))^2},
 \label{chi2}
\end{equation}
where $\xi$ represents a set of kinematic  
variables $\xi$=($\theta_{1}$,$\theta_{2}$,$\varphi_{12}$,$S$), $\Delta\sigma_{\rm st}(\xi)$ and  
$\Delta\sigma_{\rm sys}(\xi)$ denote statistical and systematic uncertainties, respectively, 
summing goes over certain set of kinematic variables and $n_{d.o.f.}$ is a number of 
data points included in this sum. The quantity defined in this way
has no precise statistical meaning, but
it still provides a measure of the description provided by different models
and allows comparison of possible deviations with experimental uncertainties.

 \begin{figure}[thb]
\includegraphics[width=8.5cm]{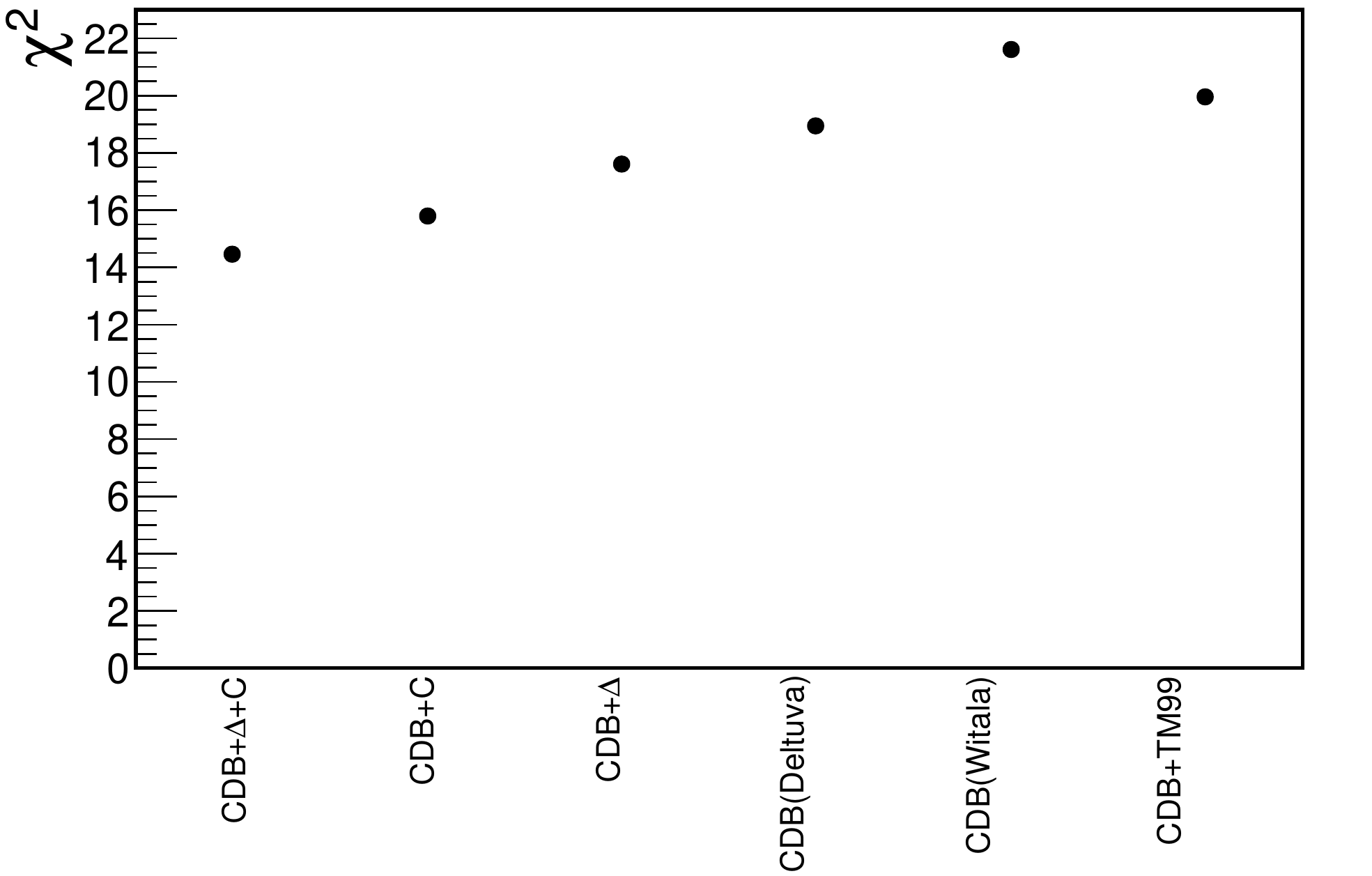}

\caption{\label{globalchi2}
(Color online) Quality of description of the differential cross section for the breakup reaction at 190\,MeV/nucleon beam energy at forward angles. 
The global $\chi^2/d.o.f.$ (systematic errors included) was obtained as a result of 
comparing the cross section data 
with each of six types of theoretical calculations specified  on the axis.
}%
\end{figure}

 \begin{figure}[thb]
\includegraphics[width=8.5cm]{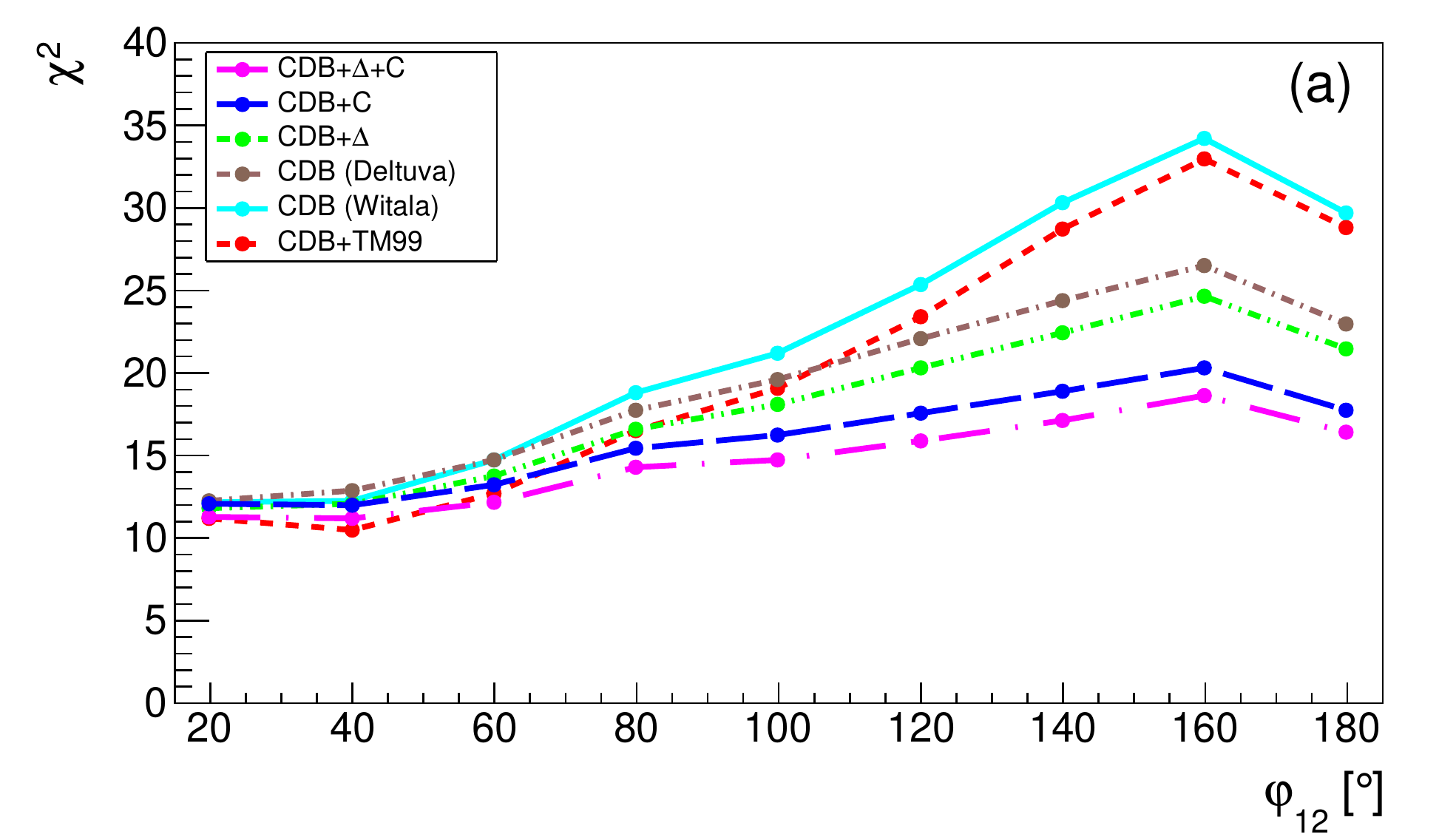}
\includegraphics[width=8.5cm]{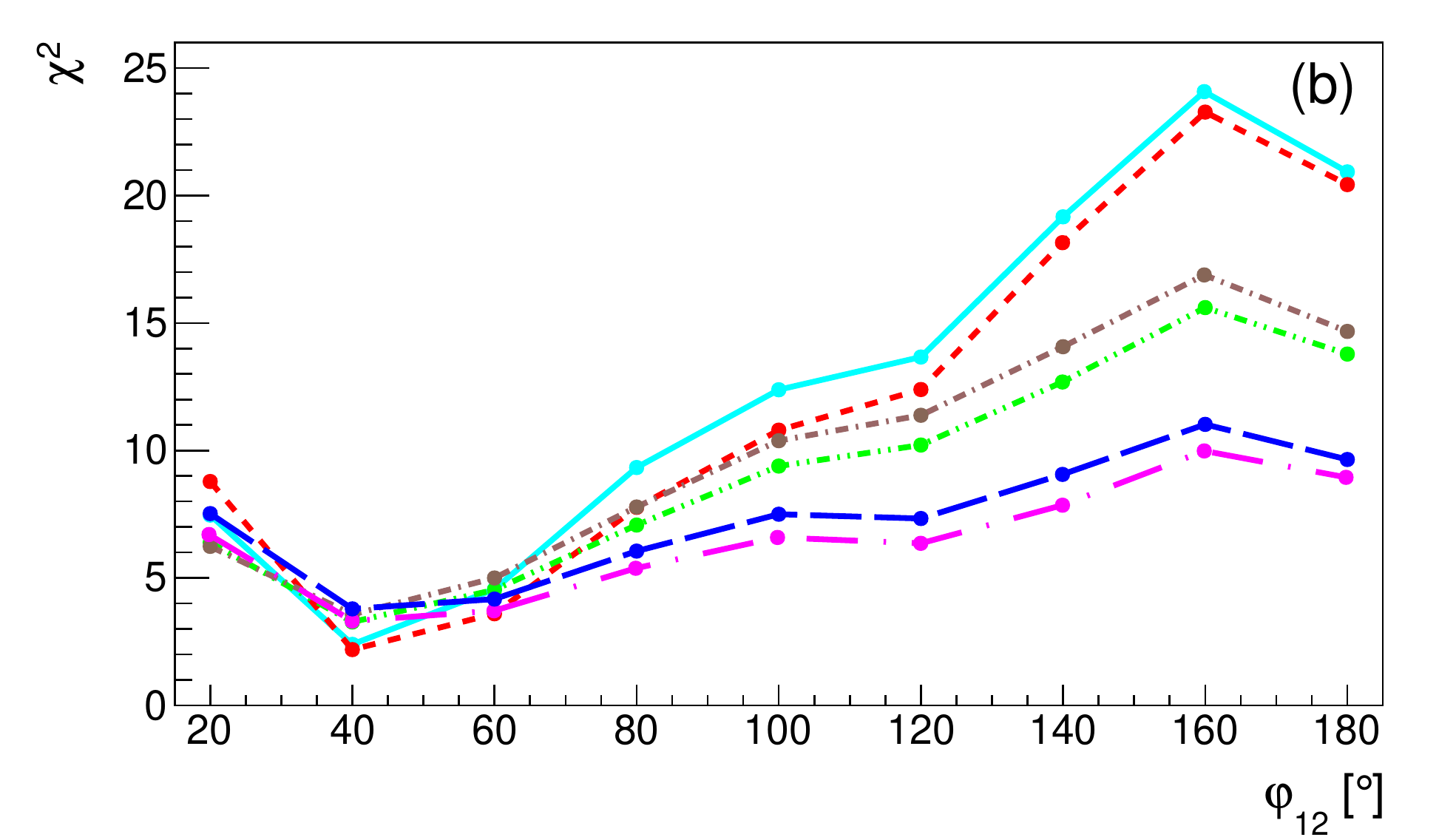}
\includegraphics[width=8.5cm]{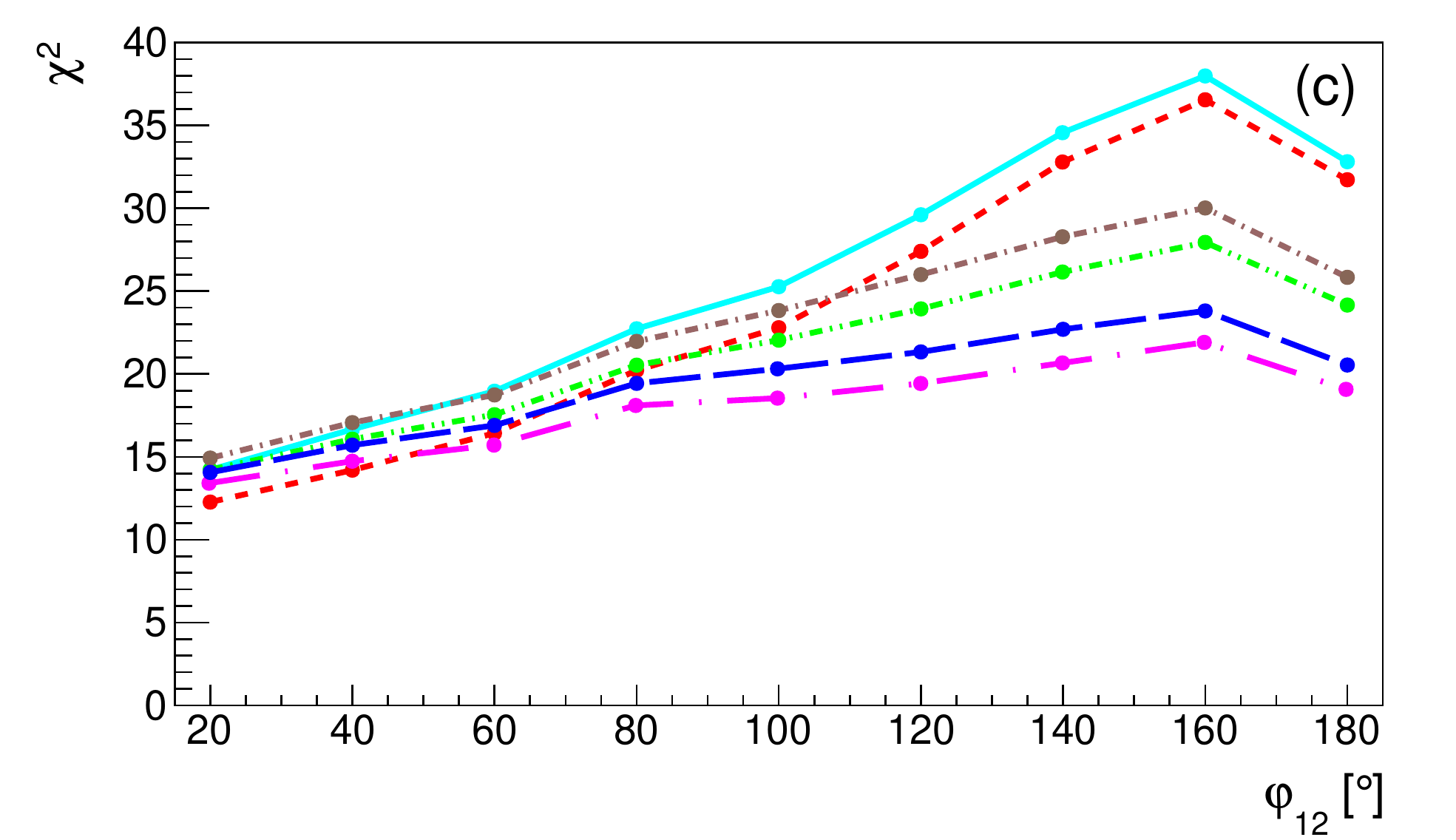}
\caption{\label{chi2_phi}
(Color online)  The $\chi^2/d.o.f.$ calculated for set of data characterised with given relative azimuthal angle $\phi_{12}$: (a) whole $S$-distrubutions were taken into account; (b) only central parts of $S$-distrubutions were taken into account; (c) only tails of $S$-distrubutions were taken into account.
}%
\end{figure}

The $\chi^2$ per degree of freedom defined above has been calculated  globally, 
individually for the kinematic configurations and, in addition, for the 
data sorted according to combination of polar angles $\theta_1, \theta_2$ or a relative 
azimuthal angle $\varphi_{12}$ of the two protons.
The global analysis (see Fig.~\ref{globalchi2}) indicates that the quality of the description of the data is worse than for the data at 170 MeV/nucleon energy, where analogously defined $\chi^2$  values covered the range between 3 and 7  \cite{WASA2020}.  Current values obtained at the beam energy of 190 MeV/nucleon are roughly 3 times higher, which is surprising given the small difference in beam energies and the same angular range in the laboratory system, thus a very similar phase space.  Global analysis indicates an improvement in description due to the inclusion of Coulomb interaction or of 3NF (TM99  force or explicit $\Delta$ isobar approach), with the best result achieved when both contributions are included. However, the aforementioned effects are small compared to the global imperfection of the description.

In order to ascertain  whether the deterioration of the description occurs in the entire phase space studied in the experiment, and to look for areas sensitive to individual contributors to the dynamics, the quality of description was studied in a function of kinematic variables. Dependence of $\chi^2$ on  $\varphi_{12}$ angle was examined first. The analysis  (see Fig.~\ref{chi2_phi}(a)) clearly indicates the region of dominance of the Coulomb effect at the largest relative azimuthal angles. Surprisingly, the improvement of description related to introducing Coulomb interactions is not so obvious  in  the region of the lowest $\phi_{12}$, close to proton-proton FSI. Calculations predict there strong  
sensitivity to Coulomb interaction, which lowers cross section, but in certain configurations data lie just between calculations with and without Coulomb interaction (see for example a configuration of $\theta _1$=5$^\circ$, $\theta _2$=5$^\circ$, $\varphi_{12}$=20$^\circ$ in  Fig.~\ref{fig5t5}), resulting in similar $\chi^2$'s for both cases.

Already at the beam energy of 170 MeV/nucleon, significant difference of shapes (widths) of cross section distributions was observed: experimental $S$ distributions were wider than theoretical ones \cite{WASA2020}. This effect might be partially attributed to relativistic effects in kinematics, but it persists even for relativistic calculations. In the data set presented here,  experimental distributions are about 12-14\% wider than theoretical ones, while variations of widths within theoretical calculations are on the level of 4\%.  To examine impact of the difference of shapes on the quality of the data description, the above presented analysis of 
$\chi^2$ in a function of $\varphi_{12}$ was repeated  separately for two data subsets: data points lying in the central part of each kinematics (maximum of the cross section distribution) and data belonging to the tails of the distributions, see respectively Fig.~\ref{chi2_phi}(b) and (c). The general trends are similar, but in the case of CDB+$\Delta$+C calculations, $\chi^2$ values are 2 or 3 times lower for peaks than for tails.
Also the contribution of tails to $\chi^2$ is rather uniform, see Fig.~\ref{chi2_phi}(c),   
while for peaks (Fig.~\ref{chi2_phi}(b)), the  quality of description in particular regions is more diverse. There is generally good description of data for $\varphi_{12}$ of 40$^{\circ}$ and 60$^{\circ}$, where the sensitivity to the interaction dynamics included in calculations is low. The Coulomb interaction has large impact at high $\varphi_{12}$. The 3NF effects, if present, practically vanish in integration over other kinematic variables. 

\begin{figure*}[thb]
\includegraphics[width=5.2cm]{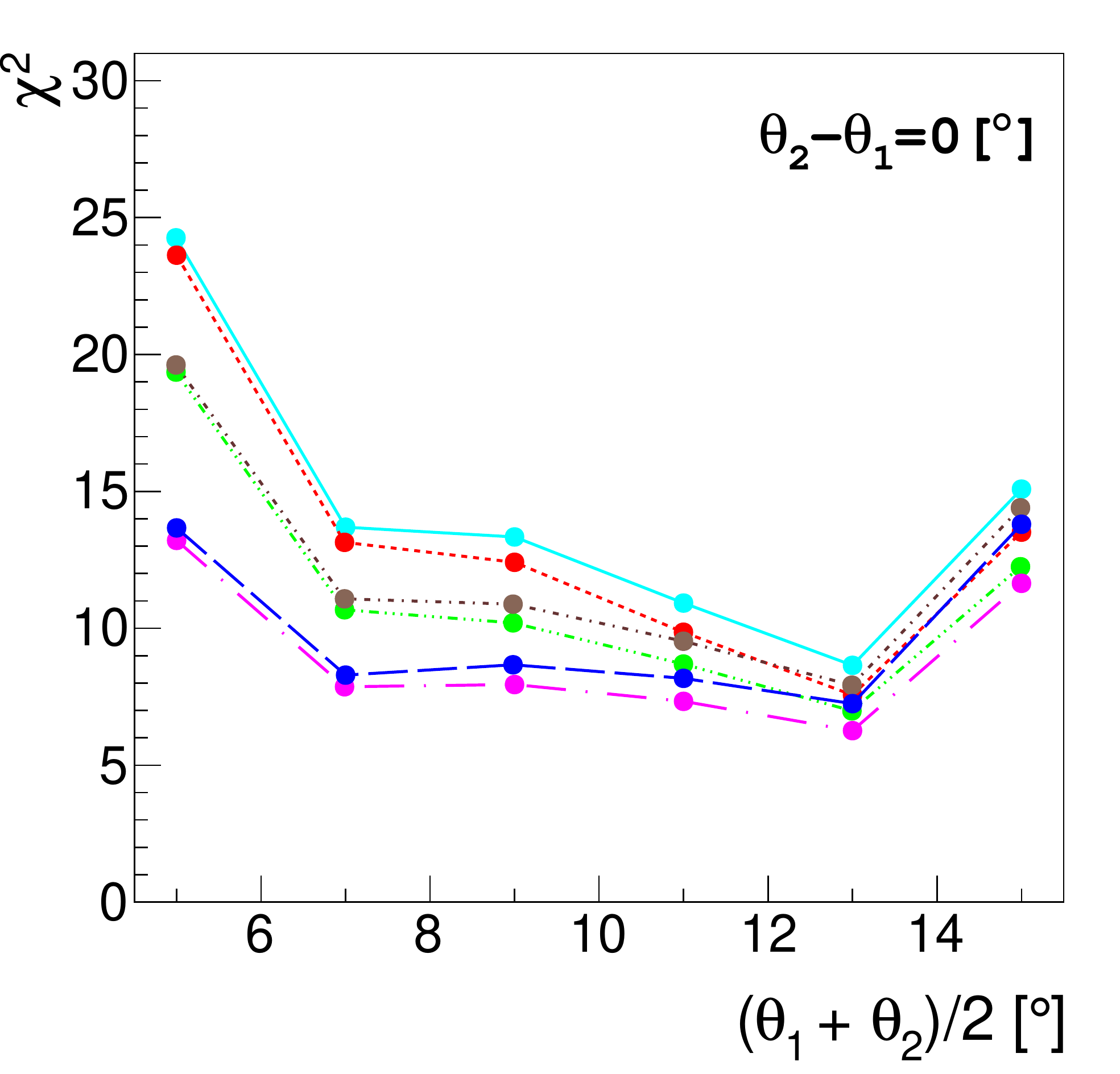}
\includegraphics[width=5.2cm]{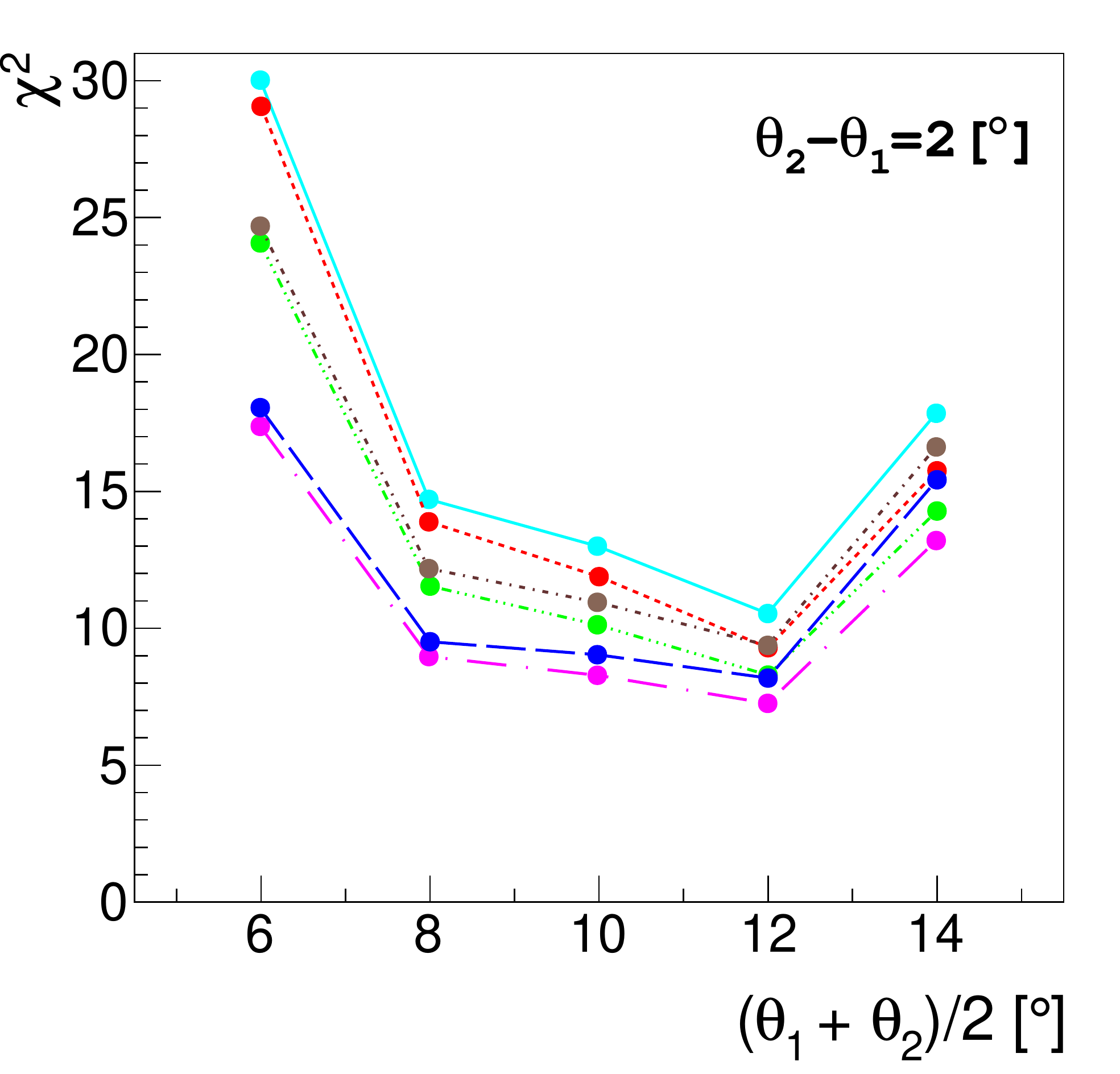}
\includegraphics[width=5.2cm]{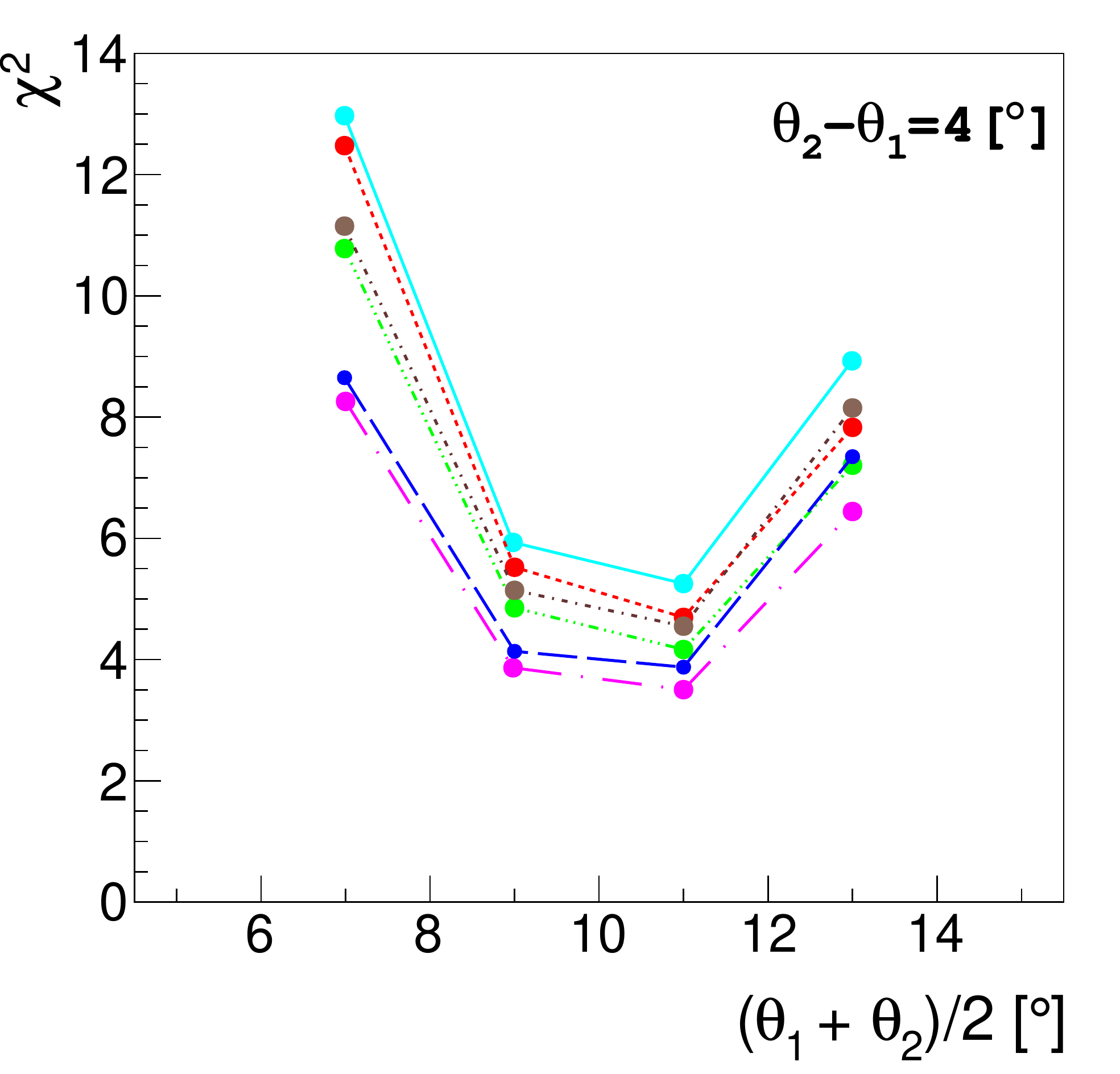}
\includegraphics[width=5.2cm]{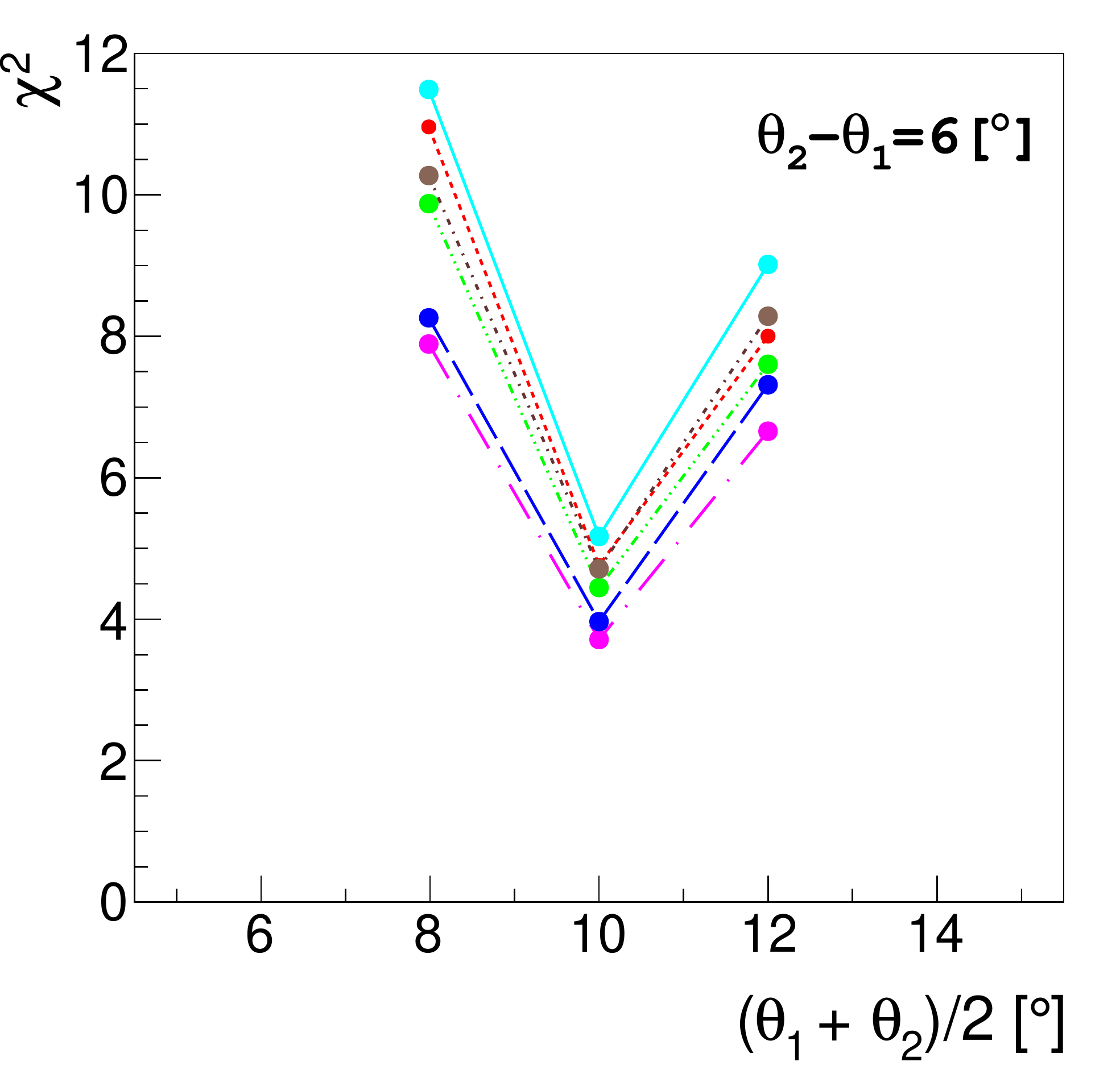}
\includegraphics[width=5.2cm]{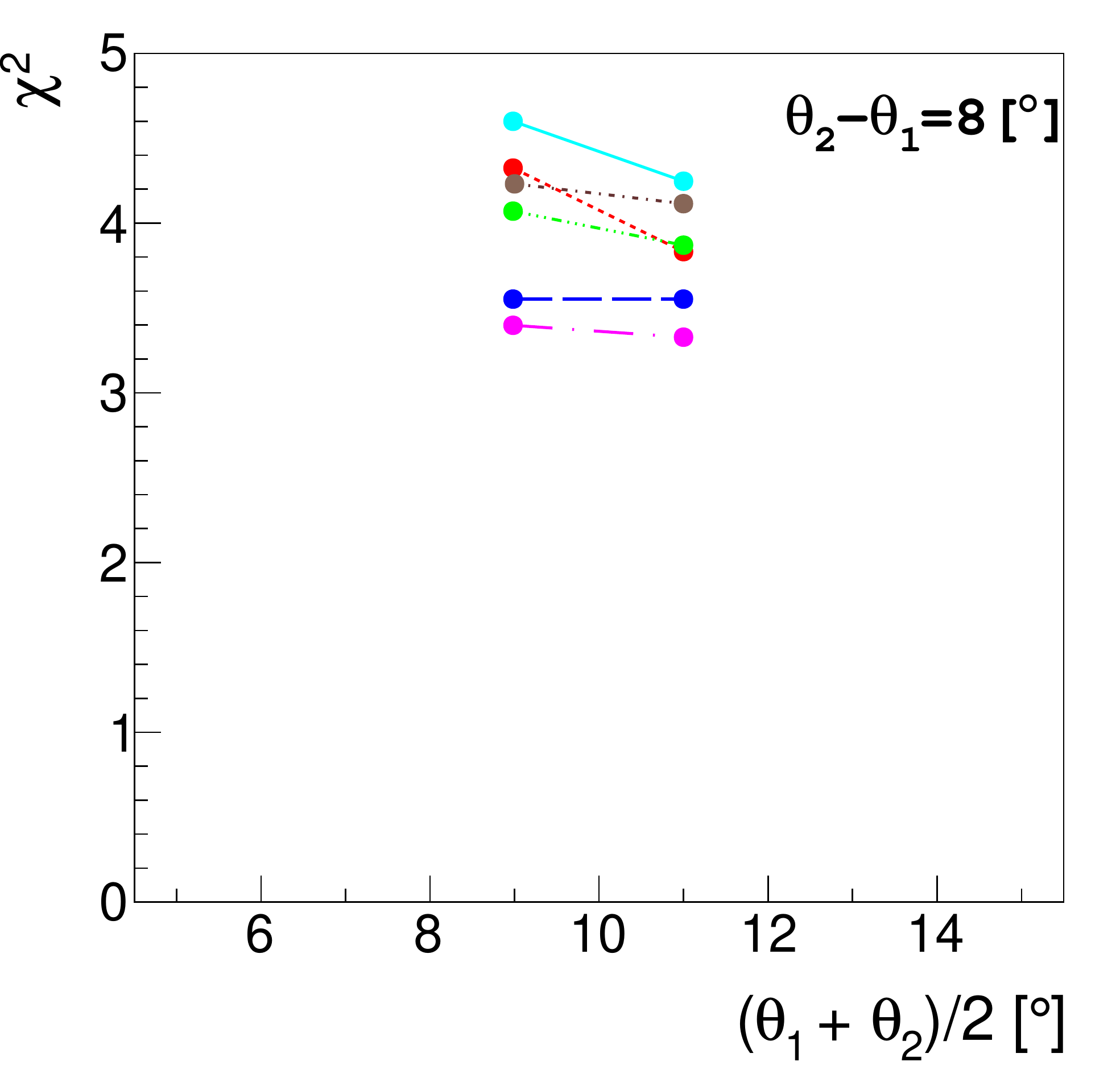}
\includegraphics[width=5.2cm]{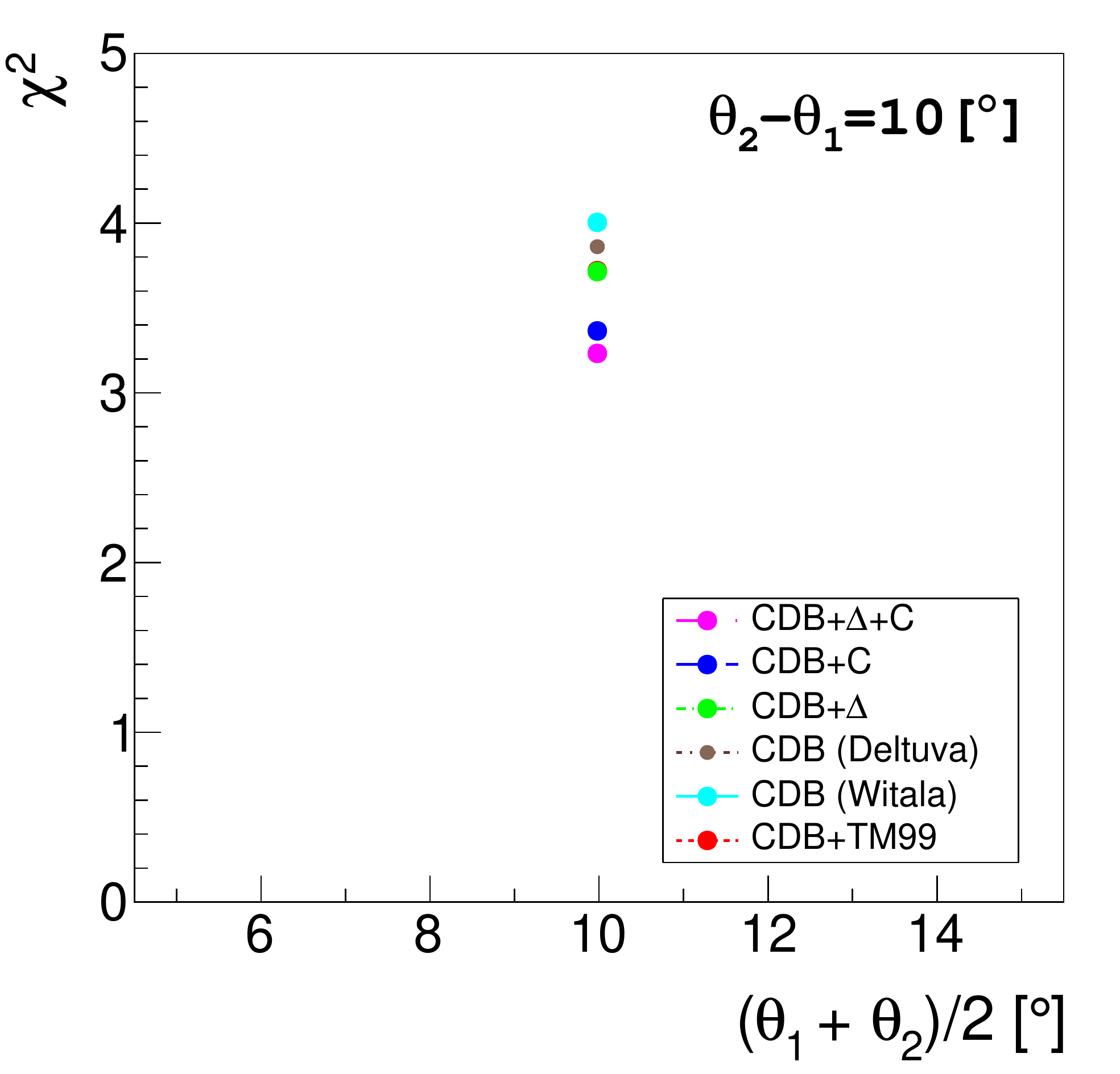}

\vspace{-0.5cm}
\caption{\label{chi2_theta}
(Color online)$\chi^2/d.o.f.$ calculated for each set of data characterised with the given combination of proton polar 
angles.  The results are ordered according to difference of polar angles 
$\theta_{12}=\theta_1-\theta_2$; in each panel results for one value of $\theta_{12}$ are 
shown. 
}%
\end{figure*}

The analysis of data sorted with respect to polar angles $\theta_1, \theta_2$ 
(Fig.~\ref{chi2_theta}) was carried out in two variables: difference $\theta_2 -\theta_1$ reflecting ``asymmetry'' of the configuration (individual panels in Fig.~\ref{chi2_theta}) and an average of the polar angles $(\theta_1+\theta_2)/2$. Clearly, 
 the Coulomb force dominate in the FSI region, characterised by  the lowest difference of polar angles and the lowest average polar angles, between 5$^\circ$ and 7$^\circ$. Moreover, description of the data improves with rising asymmetry of polar angles and/or for medium average of about 10$^\circ$.  The region of certain sensitivity to 3NF can be, in turn, identified at more symmetric configurations (three top panels) with high average  
 $(\theta_1+\theta_2)/2 \geq 13^{\circ}$, but this effect is tiny.  

\begin{figure*}[bth]
\includegraphics[width=4.19cm]{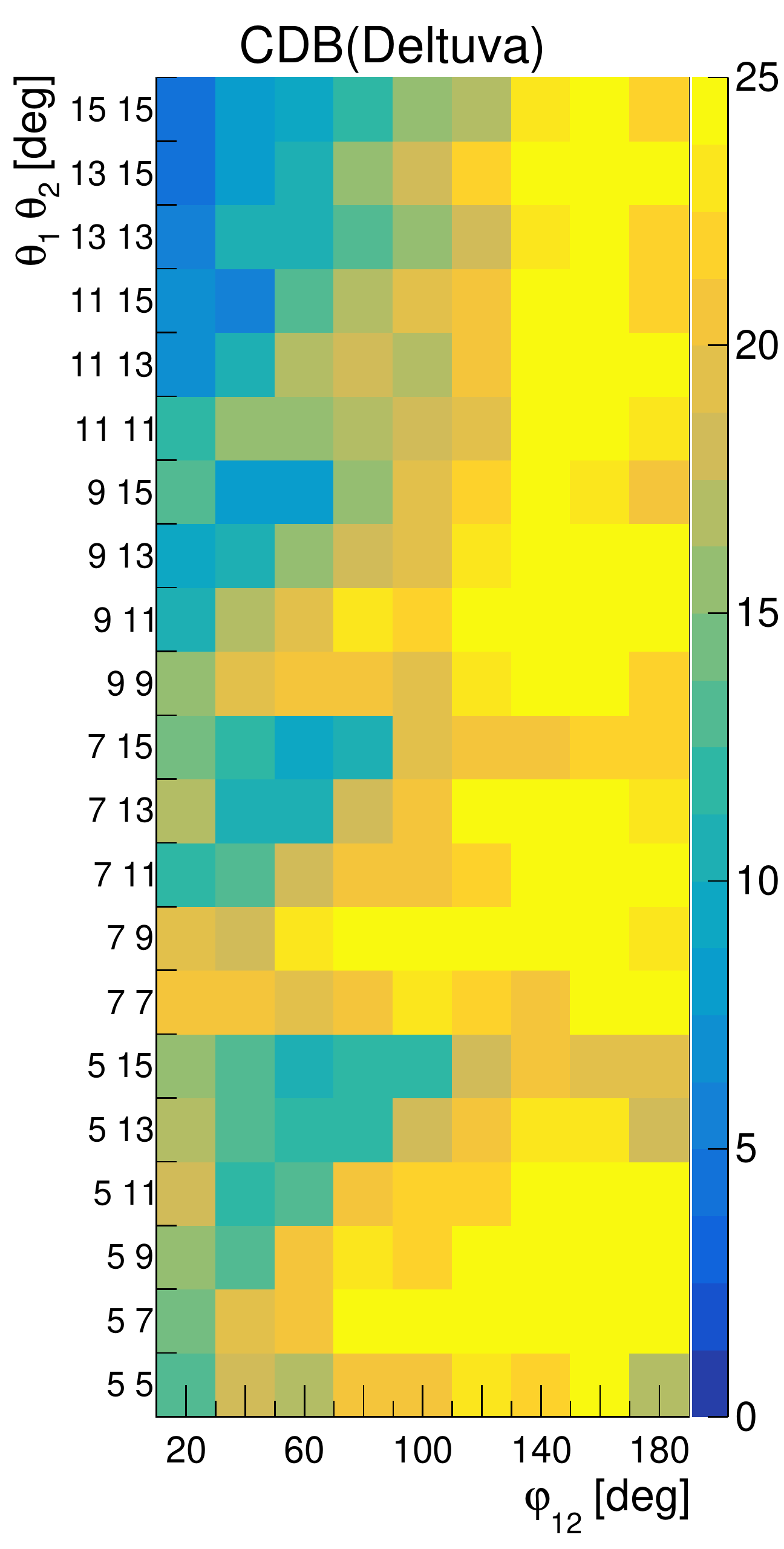}
\includegraphics[width=4.19cm]{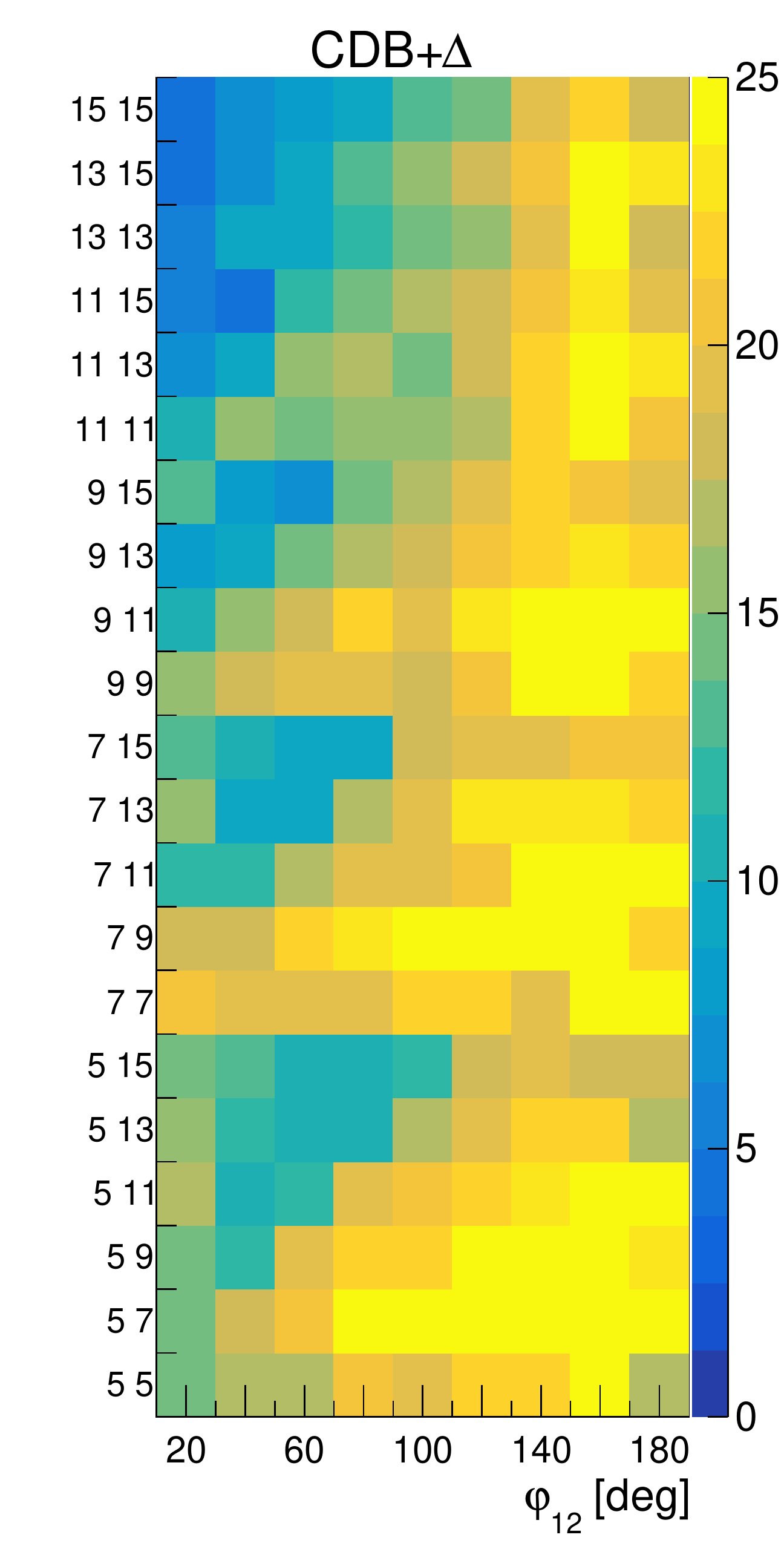}
\includegraphics[width=4.19cm]{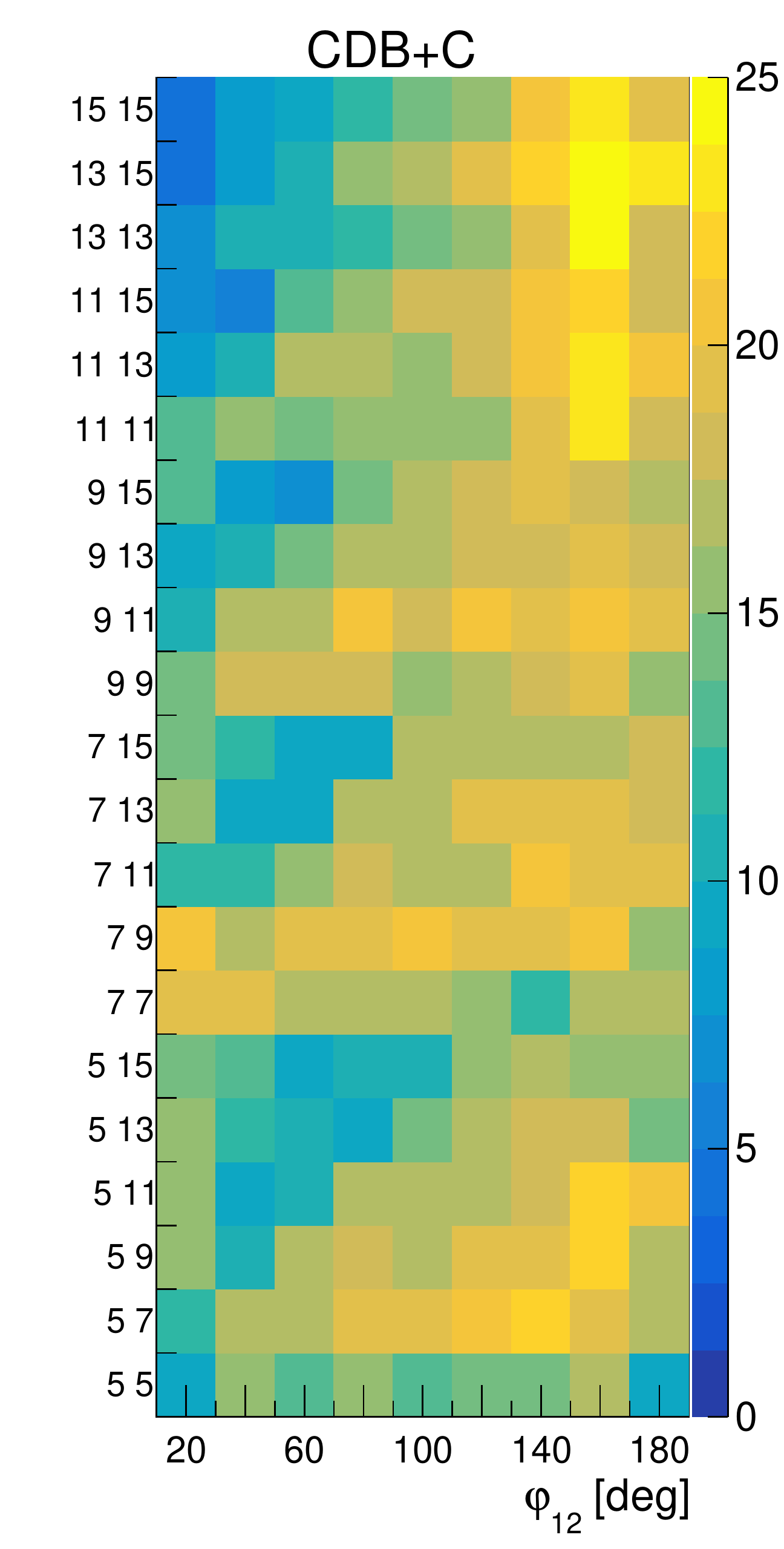}
\includegraphics[width=4.19cm]{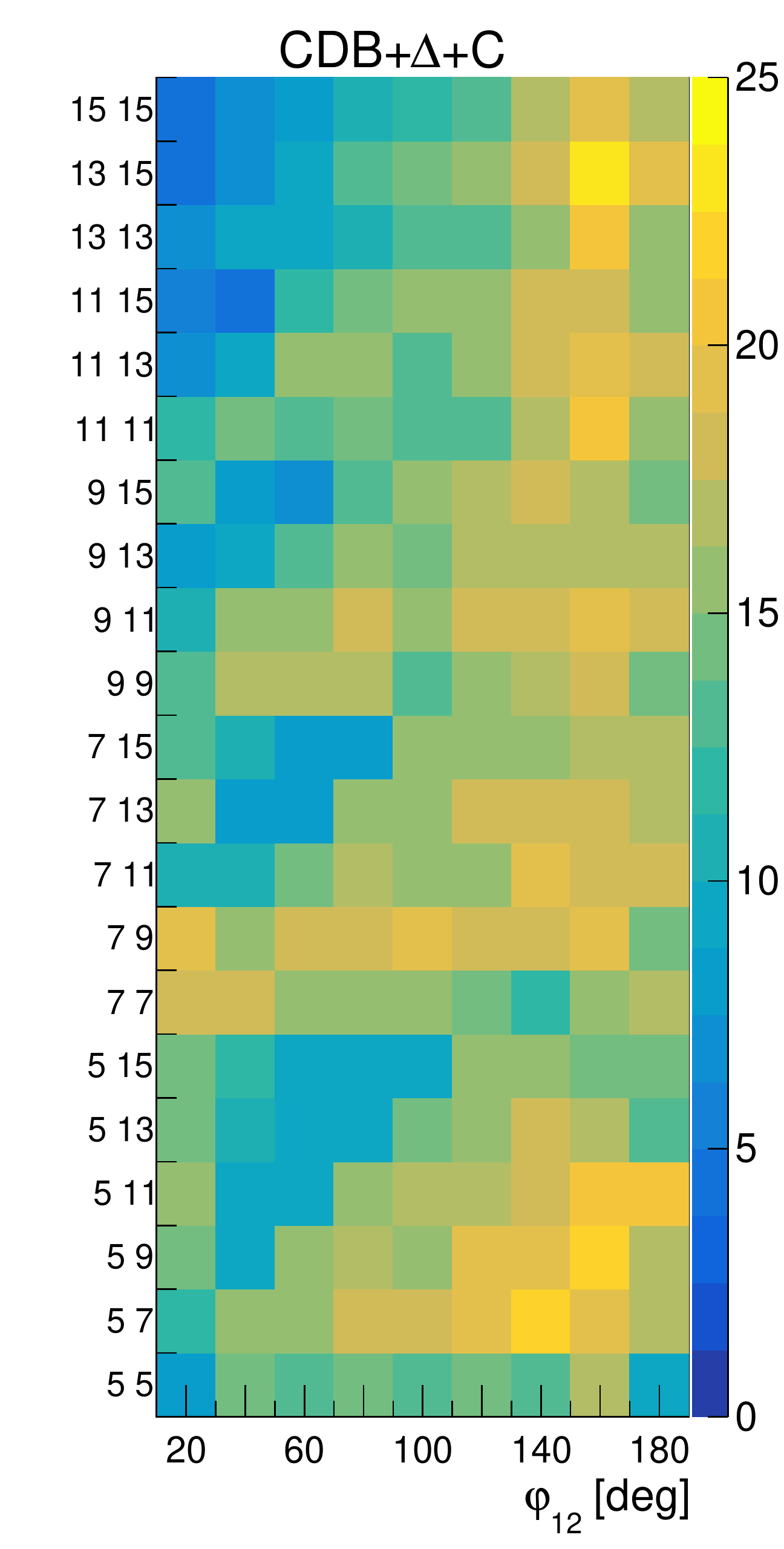}

\caption{\label{chi2maps}
(Color online) The $\chi^2/d.o.f.$ results obtained for all angular configurations separately, presented as the 2-dimensional
map  for  selected theories, specified at the top of panels}.%
\end{figure*}

 The maps of $\chi^2$ for individual configurations prepared for four theoretical calculations are shown in Fig.~\ref{chi2maps}. For the purpose of this comparison, a set of theories has been chosen that allows the Coulomb and 3NF effects to be traced both separately and together.
 For better visibility of effects, the maximum of colour z-axis was artificially set to 25. In all the maps one can observe a kind of stepped pattern corresponding to the fact, that configurations more asymmetric in polar angles are better described by the theories.  In the case of CDBonn potential alone, it is apparent that for $\varphi_{12}>100^{\circ}$ (right hand side of the map) the agreement between data and theory is poor. Only small change of picture due to 3NF (CDB+$\Delta$) is observed, as was expected from the small size of these predicted effects. On the map corresponding to CDB+C calculations the improvement of description is clearly visible. Adding both Coulomb interaction and 3NF (CDB+$\Delta$+C) leads to further small improvement of description. In the region of large polar and azimuthal angles (top right corner of the map)  a lot of room for improvement remains, as was expected from examining particular cross section distributions in Fig.~\ref{kon2rel}.   
 
\section{\label{secV}Summary}

The measurement of differential cross section of the $^{1}$H$(d,pp)n$ reaction at three beam energies, 170, 190 and 200 MeV/nucleon,  changed in a supercycle mode allowed good control over  data consistency and experimental uncertainties, constituting the basis for systematic comparison of the results. 
In this work, the differential cross section for beam energy 190 MeV/nucleon is presented for 189  configurations characterized with forward proton emission angles in the laboratory system and compared to a set of theoretical calculations. In this region,  predicted 3NF effects are small, while calculated  Coulomb effects are significant, especially at low polar angles of emitted protons. Relativistic calculations were  performed for several selected configurations with CD-Bonn potential alone, showing significant influence on the cross section compared to the same calculations in non-relativistic approach. 

Confronting these prediction with the data, we clearly confirm observation of the Coulomb effects, though in some cases calculations including Coulomb interactions underestimate the data. It is not clear, whether this can be attributed to relativistic effects or other missing dynamical effect.  Moreover, general deterioration of the overall data description is observed, as compared to data collected at slightly lower beam energy of 170 MeV/nucleon.  Most data lies above the theoretical predictions, but in many cases this difference is contained within systematic uncertainties, while the significant disagreements are observed in tails of the cross section $S$-distributions. This problem is related to different widths of the theoretical and experimental distributions. Apart of this issue, the description of data in the peaks is quite successful for majority of configurations, in particular after Coulomb interaction is included.

We have, however, localized  a set of kinematical configurations of large disagreements between the experimental data  and all the available theoretical calculations. This mainly  concerns configurations with polar angles of $\theta_{1}=\theta_{2}=  13^{\circ}$ or $15^{\circ} $ and  coplanar or nearly coplanar geometries ($\varphi _{12} =160^\circ, 180^\circ$), where  the differential cross section is the lowest in the studied phase space. The  discrepancies between data and theories roughly of the factor of 2 are observed.  
The latter finding is consistent with that followed from the previously analyzed cross section data at  170 MeV/nucleon. There,  local underestimation of data with theories was observed for exactly the same configurations. Comparisons of these two data sets indicate the discrepancy rising  with the beam energy. Calculations performed in the relativistic approach,  limited to the CD-Bonn potential, further increase the discrepancy, leaving room for very large effects of the 3NF, well beyond the predictions of current models. This suggests either problem with currently available 3NF's at so high energy, or huge relativistic effects in the 3NF's. 

Importantly, similar underestimation the data by theoretical predictions was observed in the cross section data collected at the same center of mass energy, but in the direct kinematics $^{2}$H$(p,pp)n$ \cite{Mardanpour_PhD, Mardanpour08, Hajar2021}. There,  the discrepancies were also observed locally, in symmetric configurations with   $\varphi_{12}\geq160^{\circ}$.  In spite of these  similarities, those configurations corresponded to a different $\theta_{CM}$ region. Also,  in the regions of disagreements the predicted relativistic effects were  small: either negligible, or  shifting the results of calculations slightly towards the data, though far too small to solve the problem.  

Systematic studies of the deuteron breakup in collision with proton at relatively high energies are important for testing current approaches to describe nuclear interactions. The differential cross section exhibits surprising sensitivity to missing elements of the dynamics, also in the regions where current 3NF models do not predict any spectacular effects.  Since relativistic
calculations based on pure $NN$ interaction show the effect
opposite to the one needed for the correct data description, an approach that takes into account relativistic effects together with a full complexity of interaction potential, including 3NF, would be of great interest. 

\begin{acknowledgments}
This work was partly supported by the National Science Centre, Poland, under Grant No. IMPRESS-U 2024/06/Y/ST2/00135 and
by the Excellence Initiative–Research University Program at the Jagiellonian University in Kraków. The numerical calculations
were partly performed on the supercomputers of the JSC, Jülich, Germany. 
We thank the COSY crew for their work and the excellent conditions during the beam time.

\end{acknowledgments}

\appendix

\section{}

Appendix A~contains all the differential cross section distributions 
for the $^{1}$H$(d,pp)n$ breakup reaction at the beam energy of 190~MeV/nucleon obtained in the analysis described in this paper. 
Each set presents 9 configurations corresponding to individual azimuthal angles $\varphi_{12}$ from $20^{\circ}$ to $180^{\circ}$ and one combination of $\theta_{1}$ and $\theta_{2}$ specified in the Figure. Since the 3NF effects are generally small and the results weakly depend on the model used, for clarity only calculations with one 2N potential (CD-Bonn) and one 3NF model (explicit treatment of $\Delta$ isobar) are shown.  

 \begin{center}
\begin{figure*}[t]
\subfigure{
\includegraphics[width=16.5cm]{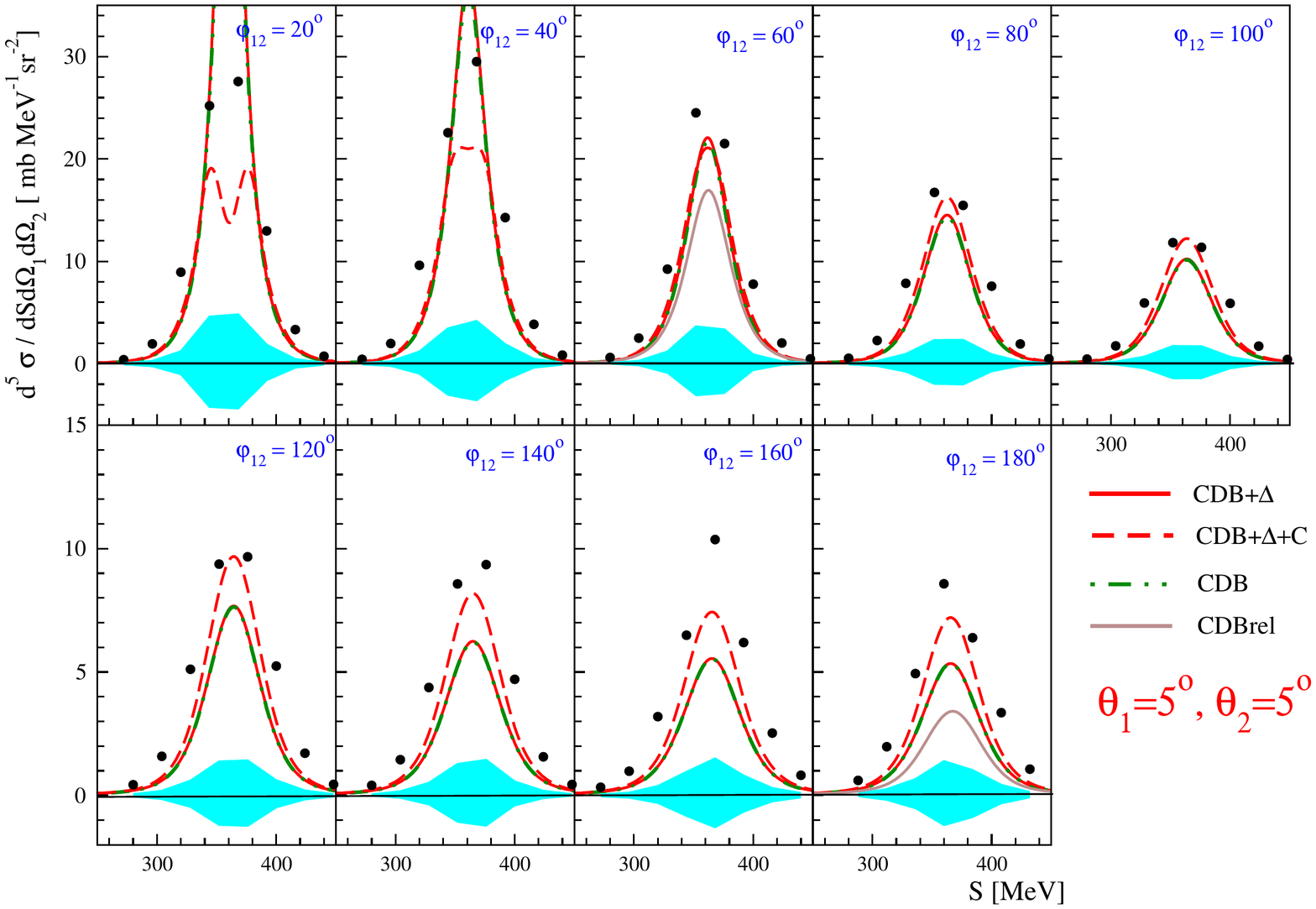}
}
\vspace*{-10mm}
\subfigure{
\includegraphics[width=16.5cm]{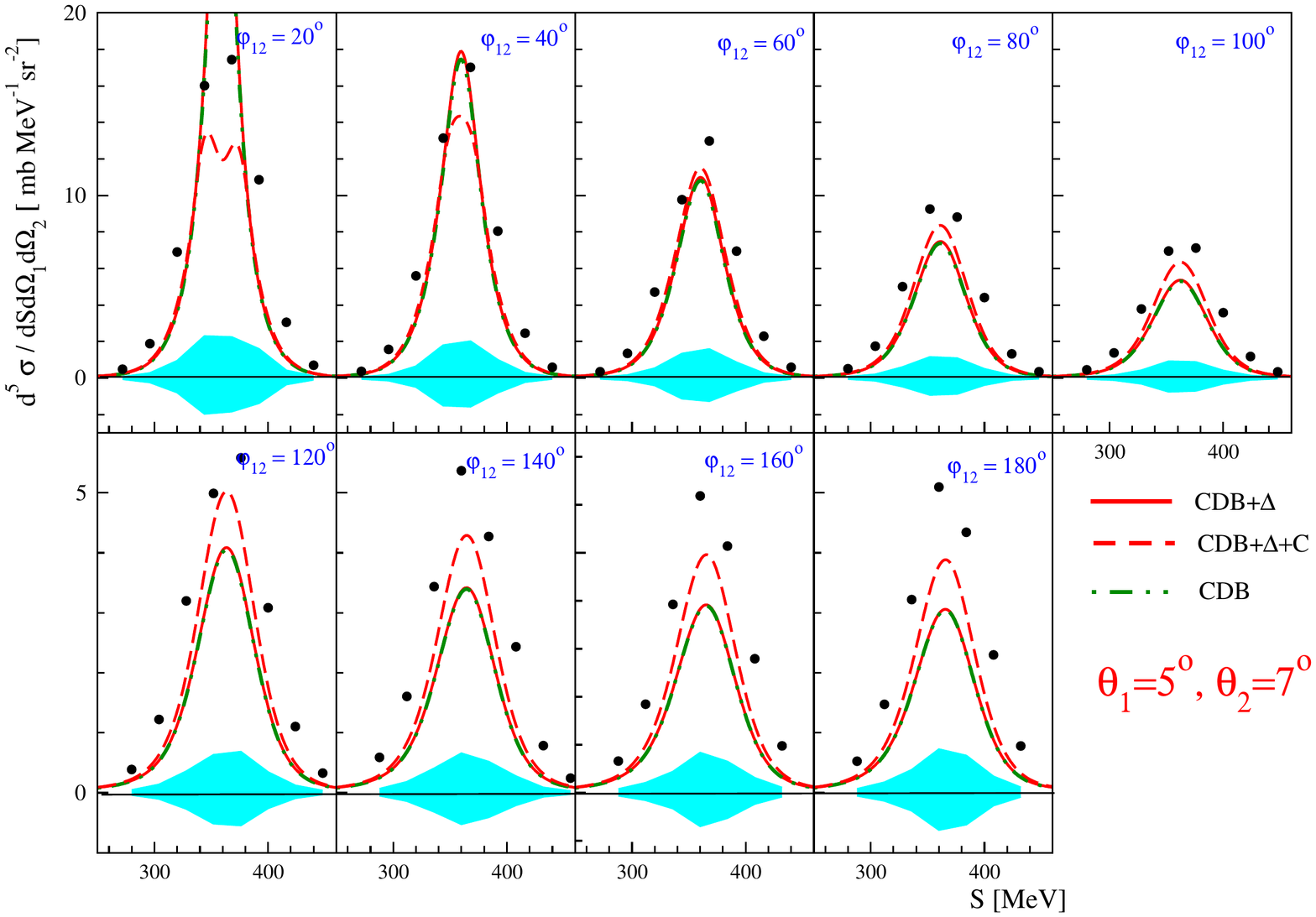}
}

\caption{(Color online) Differential cross section of $^{1}$H$(d,pp)n$ breakup reaction at beam energy of 190\,MeV/nucleon shown in function of $S$ variable. The presented data belong to 9 kinematic configurations characterised with the same combinations of proton polar angles ($\theta _1$=5$^\circ$, $\theta _2$=5$^\circ$ in the top part and $\theta _1$=5$^\circ$, $\theta _2$=7$^\circ$ in the bottom part) and various relative azimuthal angles of the two protons, indicated in the individual panels. Statistical errors are smaller than the size of the points. Systematic uncertainties are represented by bands. Theoretical calculations are shown as lines specified in the legend.
\label{fig5t5}}
\end{figure*}
\end{center}  

\begin{center}
\begin{figure*}[h]
\subfigure{
\includegraphics[width=16.5cm]{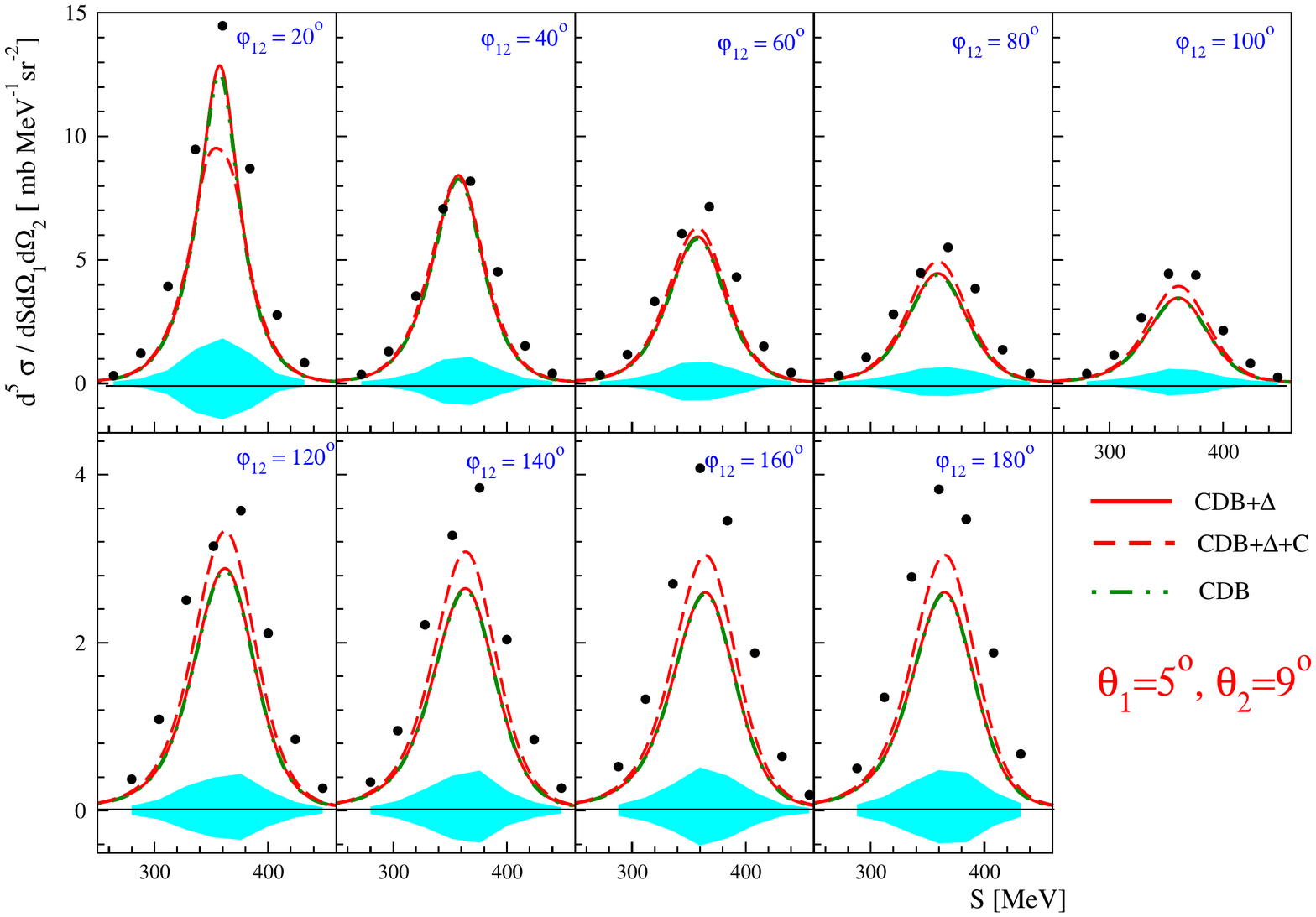}
}
\subfigure{
\includegraphics[width=16.5cm]{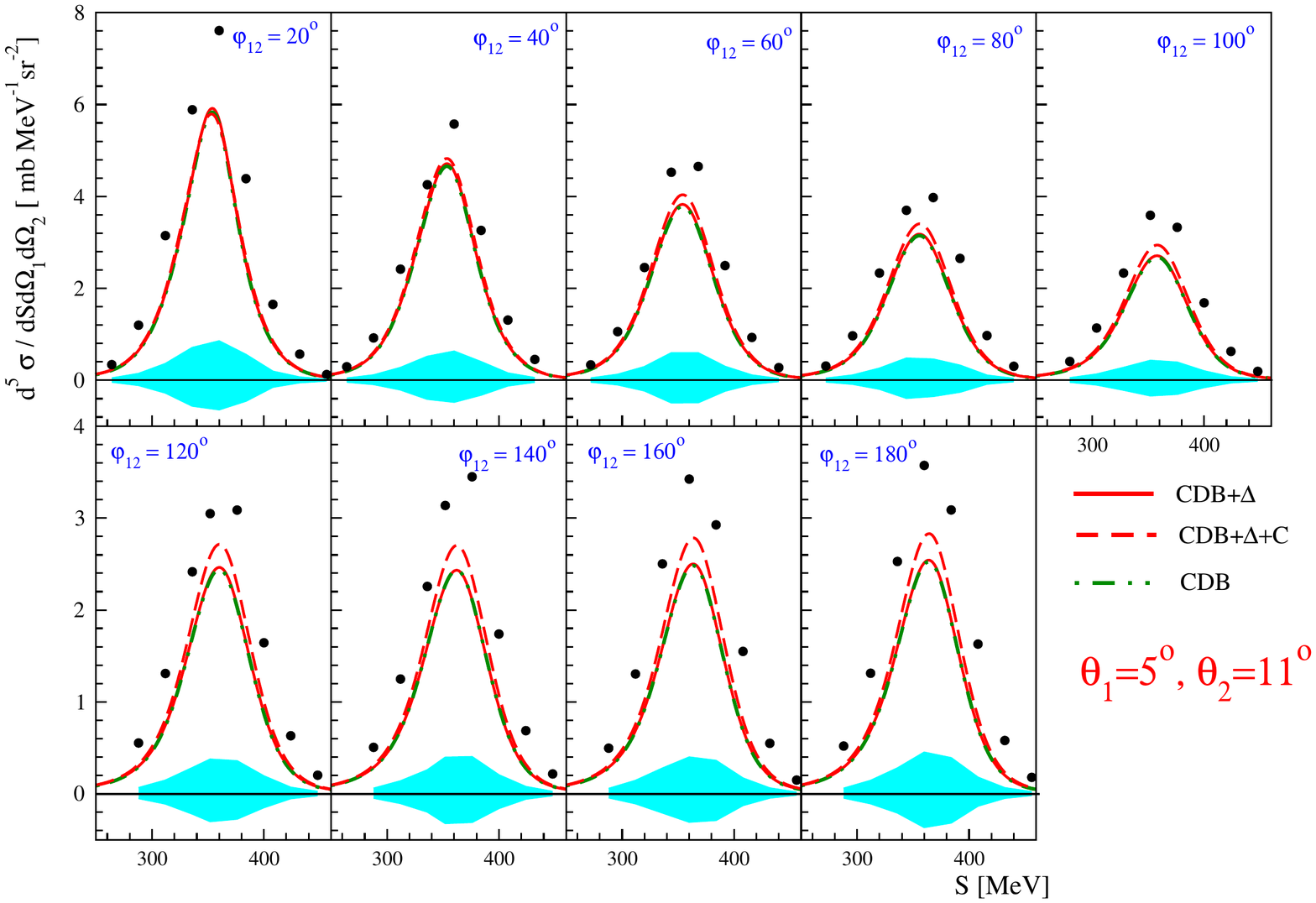}
}
\caption{(Color online) The same as Fig.~\ref{fig5t5} but for kinematic configurations with the proton polar angles  $\theta _1$=5$^\circ$, $\theta _2$=9$^\circ$ and $\theta _1$=5$^\circ$, $\theta _2$=11$^\circ$. 
\label{fig5t19}}
\end{figure*}
\end{center} 

\begin{center}
\begin{figure*}[h]
\subfigure{
\includegraphics[width=16.5cm]{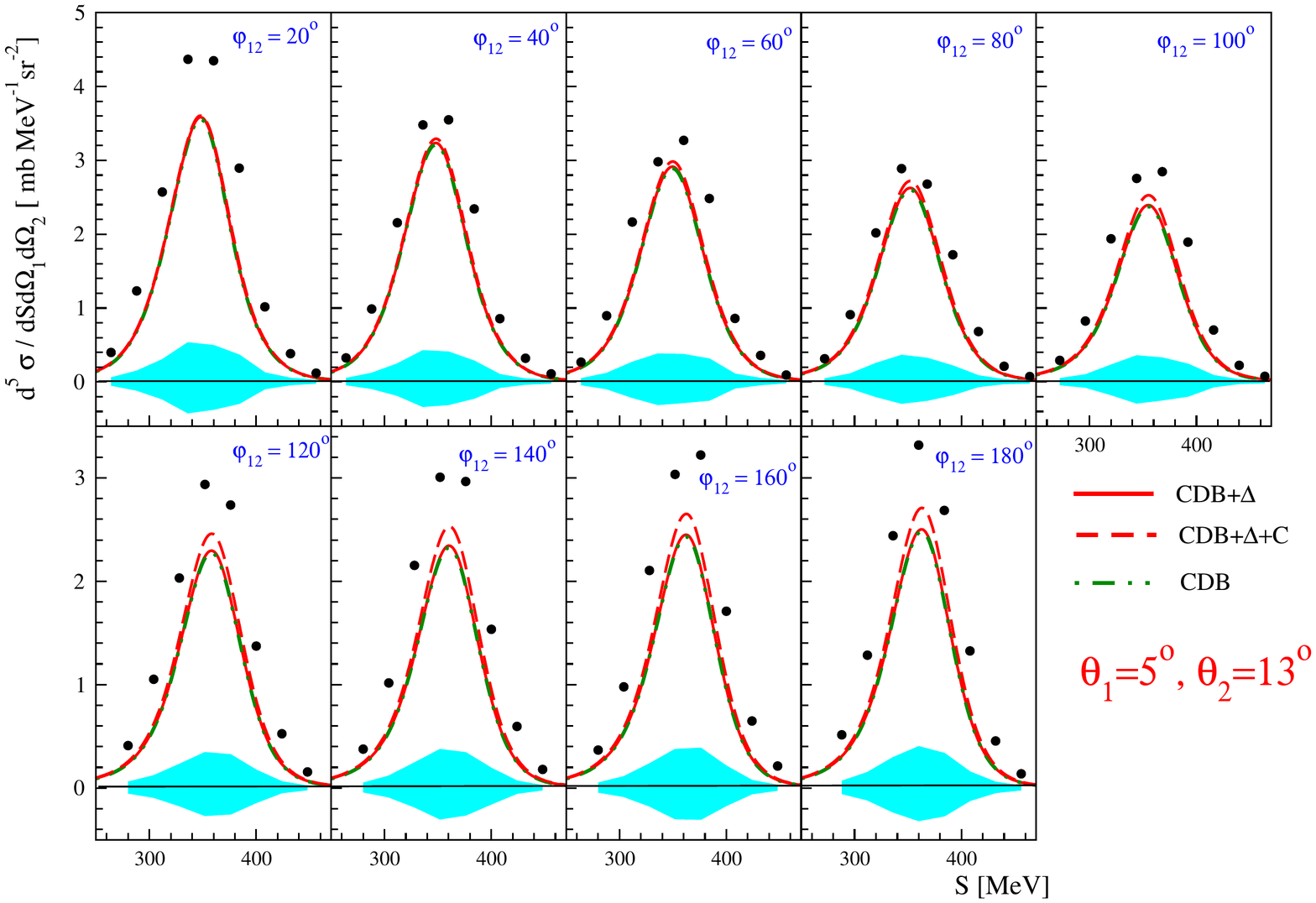}
}
\subfigure{
\includegraphics[width=16.5cm]{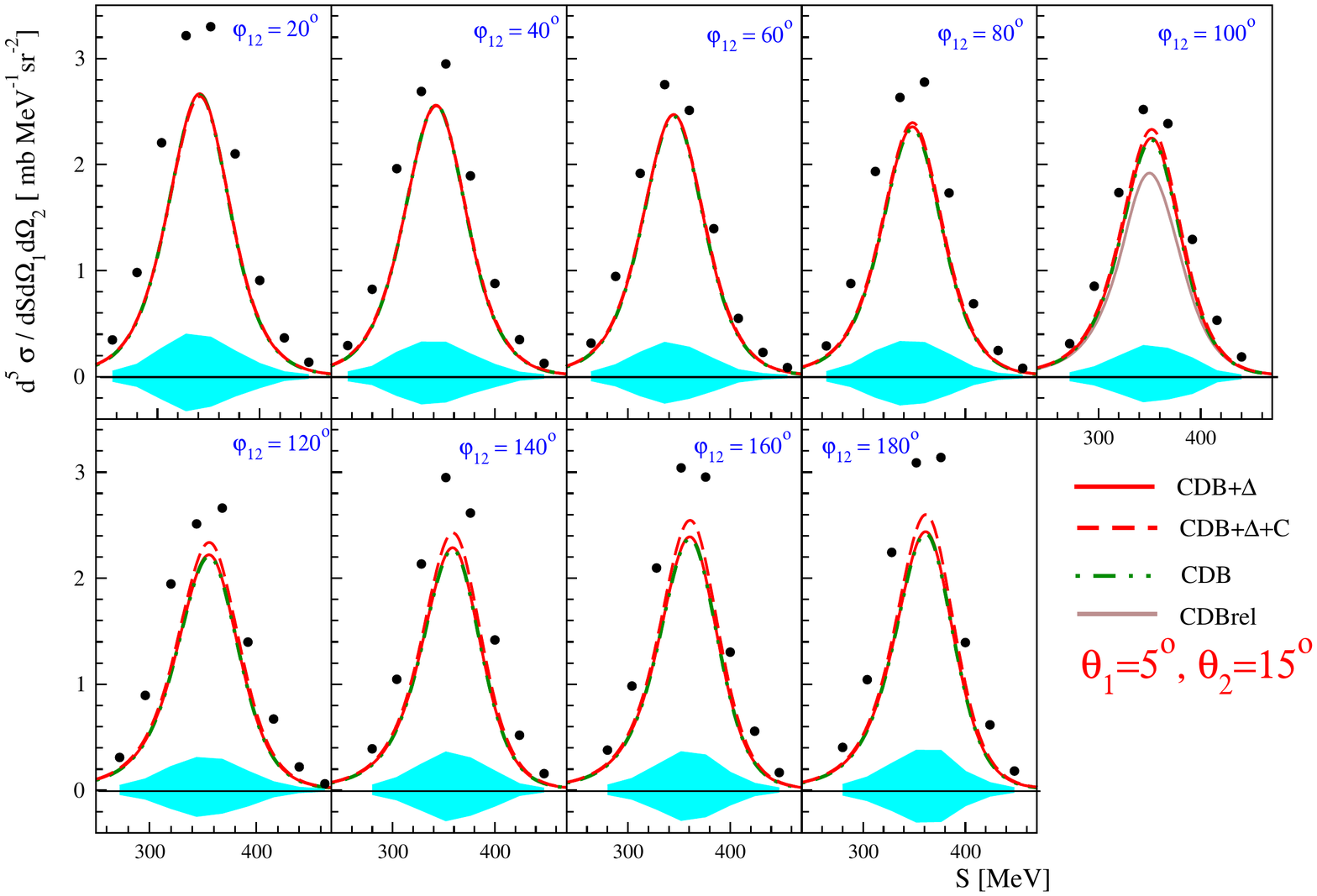}
}
\caption{(Color online) The same as Fig.~\ref{fig5t5} but for kinematic configurations with the proton polar angles  $\theta _1$=5$^\circ$, $\theta _2$=13$^\circ$ and $\theta _1$=5$^\circ$, $\theta _2$=15$^\circ$. 
\label{fig5t13}}
\end{figure*}
\end{center}  

\begin{center}
\begin{figure*}[h]
\subfigure{
\includegraphics[width=16.5cm]{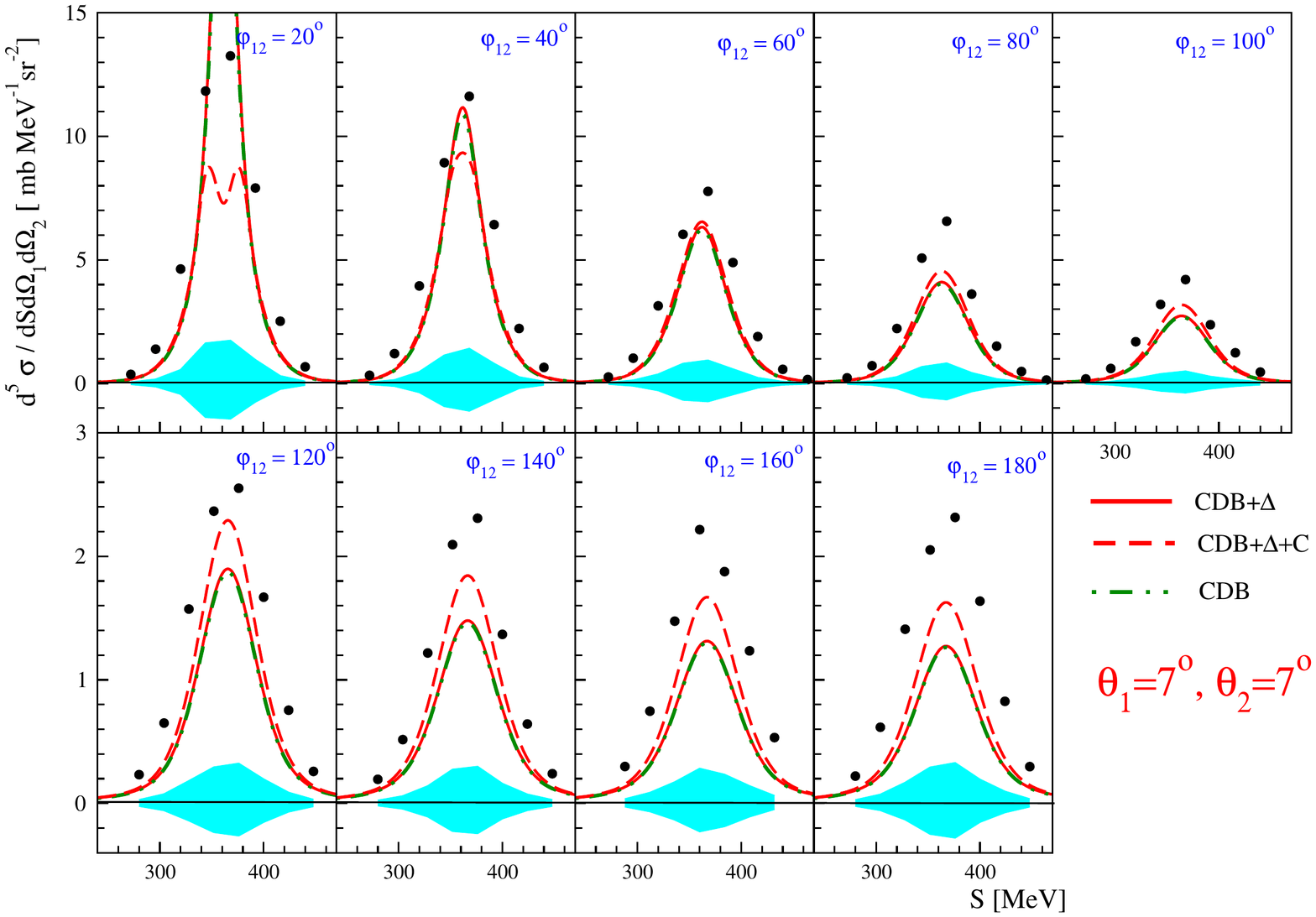}
}
\subfigure{
\includegraphics[width=16.5cm]{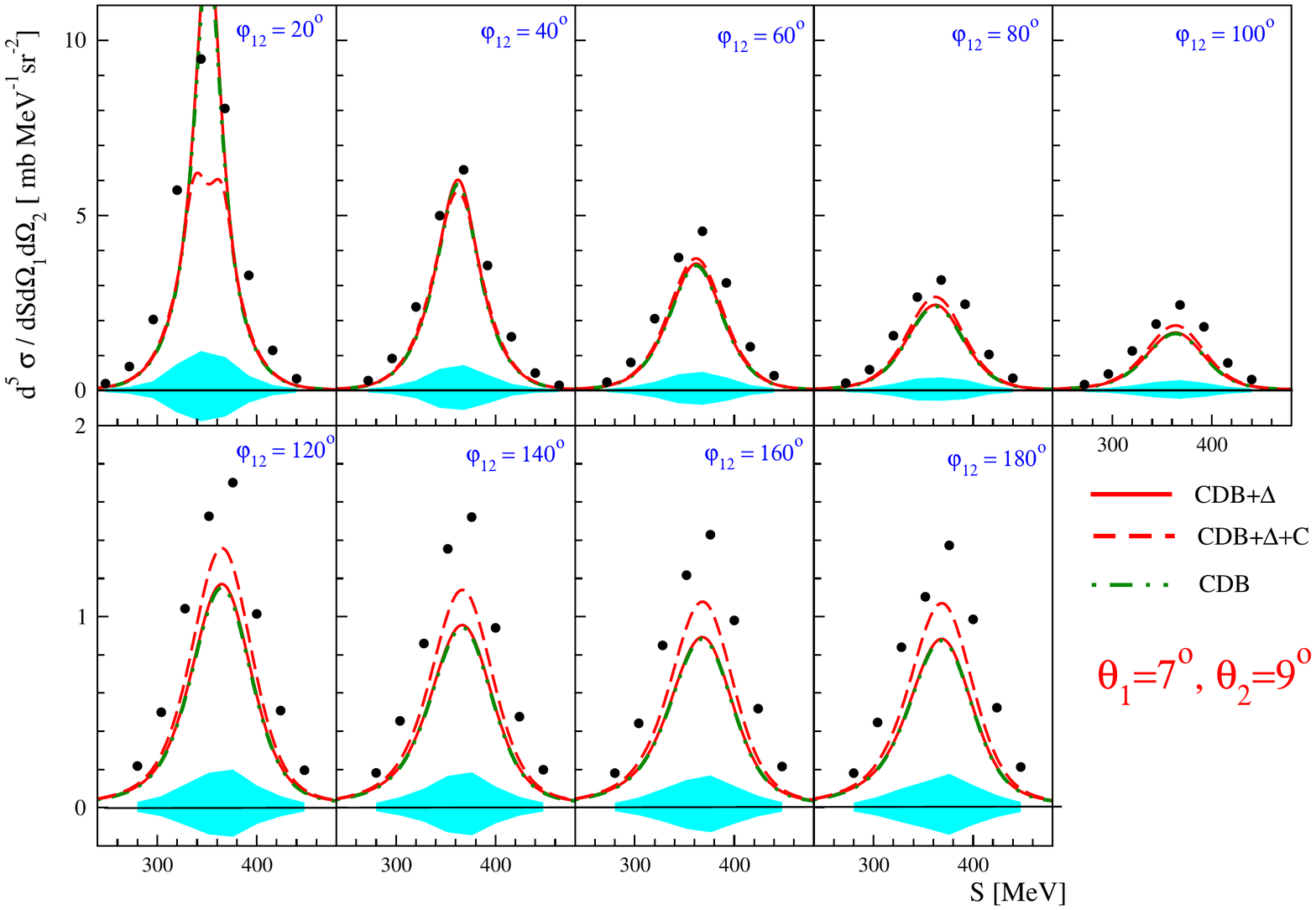}
}
\caption{(Color online) The same as Fig.~\ref{fig5t5} but for kinematic configurations with the proton polar angles  $\theta _1$=7$^\circ$, $\theta _2$ = 7$^\circ$ and $\theta _1$=7$^\circ$, $\theta _2$=9$^\circ$. 
\label{fig7t7}}
\end{figure*}
\end{center}

\begin{center}
\begin{figure*}[h]
\subfigure{
\includegraphics[width=16.5cm]{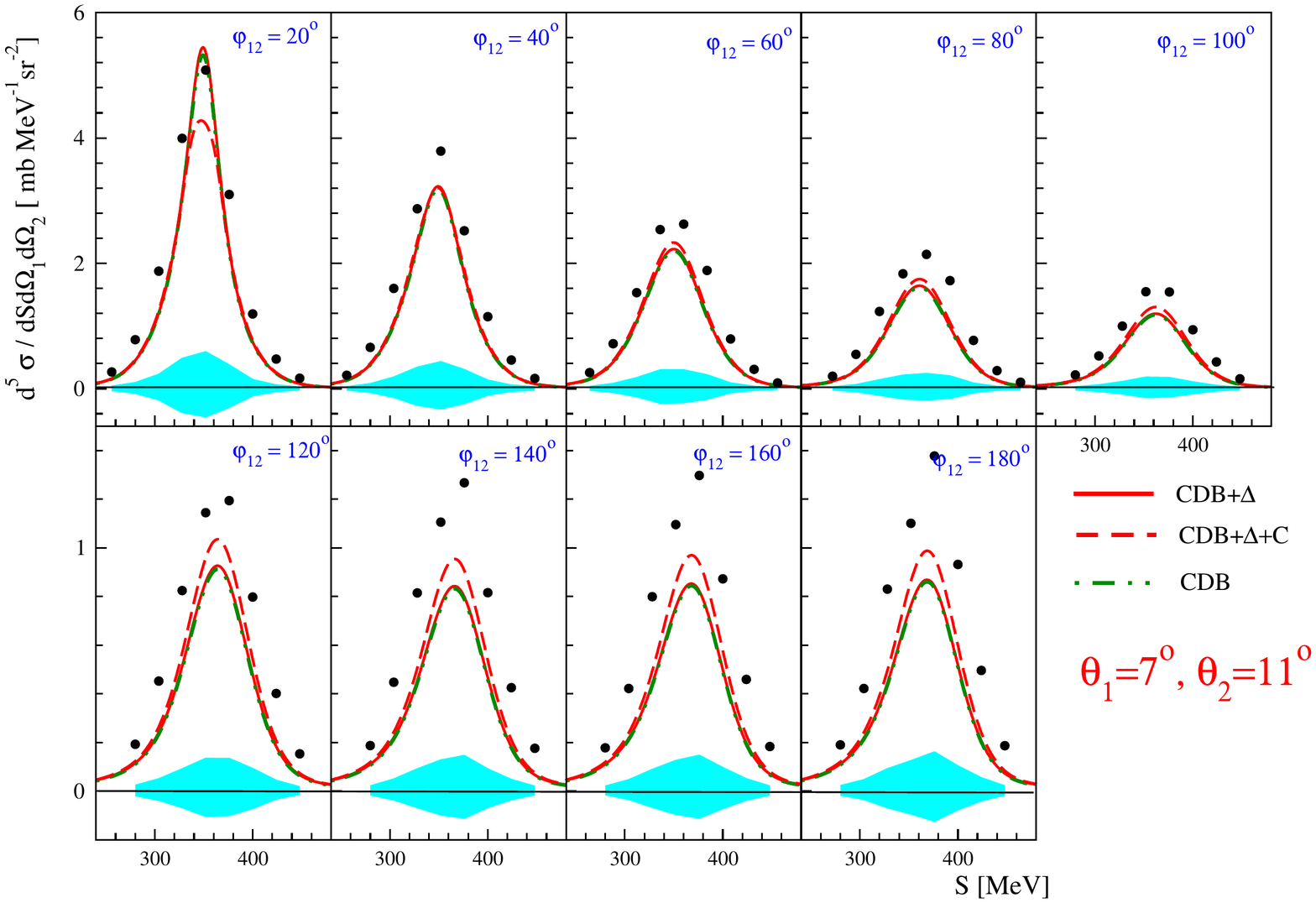}
}
\subfigure{
\includegraphics[width=16.5cm]{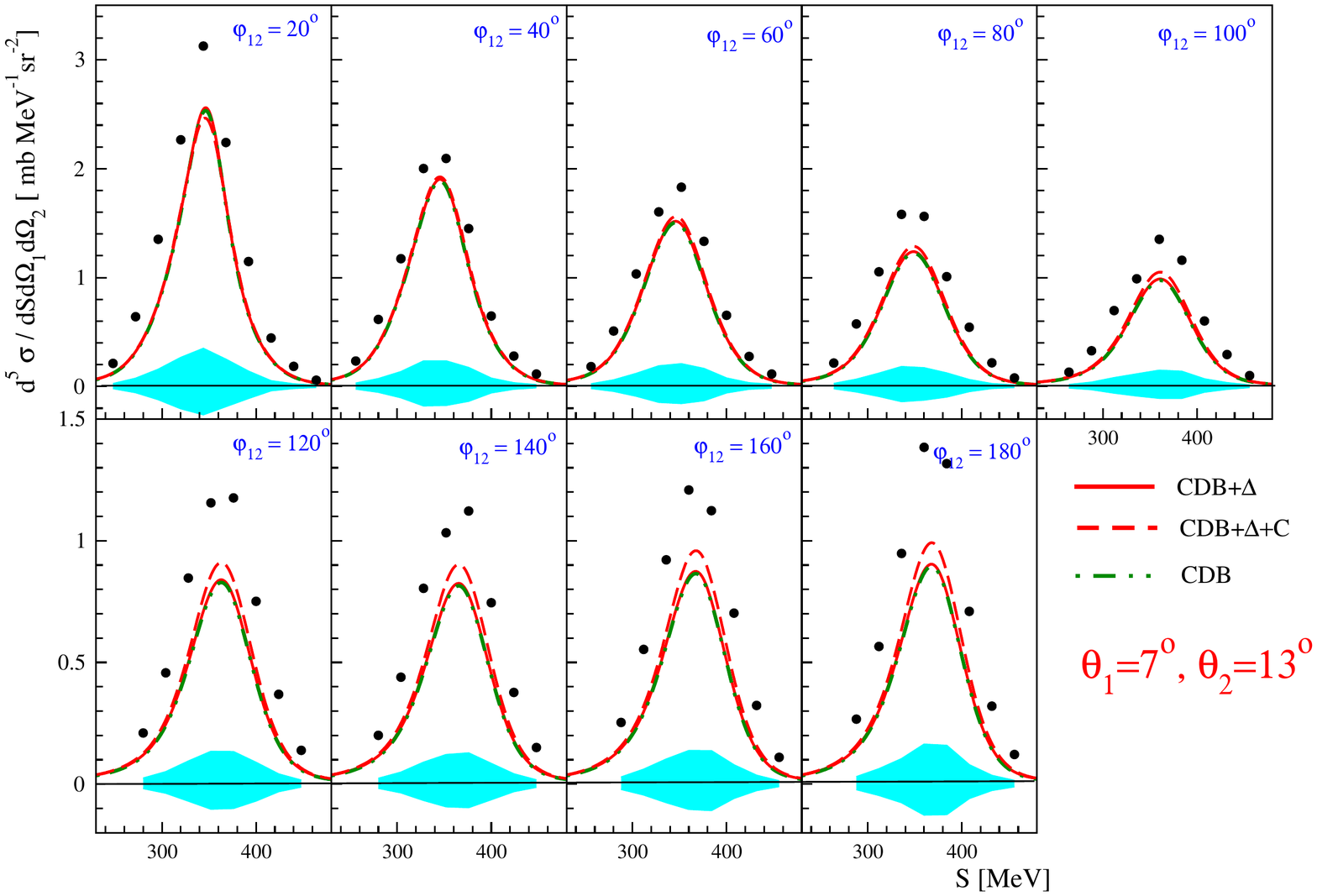}
}
\caption{(Color online) The same as Fig.~\ref{fig5t5} but for kinematic configurations with the proton polar angles  $\theta _1$=7$^\circ$, $\theta _2$=11$^\circ$ and $\theta _1$=7$^\circ$, $\theta _2$=13$^\circ$. 
\label{fig7t11}}
\end{figure*}
\end{center}  

\begin{center}
\begin{figure*}[h]
\subfigure{
\includegraphics[width=16.5cm]{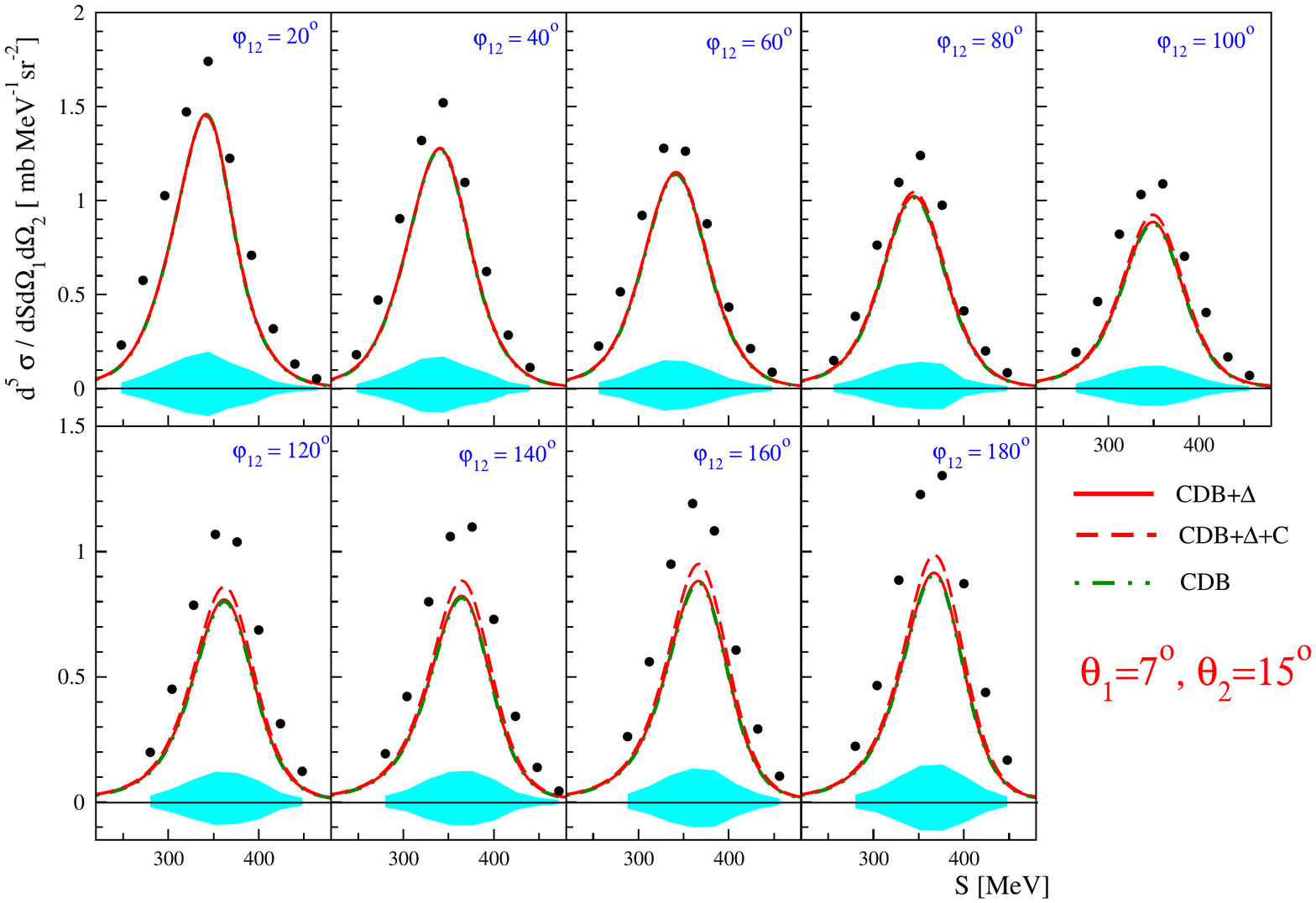}
}
\subfigure{
\includegraphics[width=16.5cm]{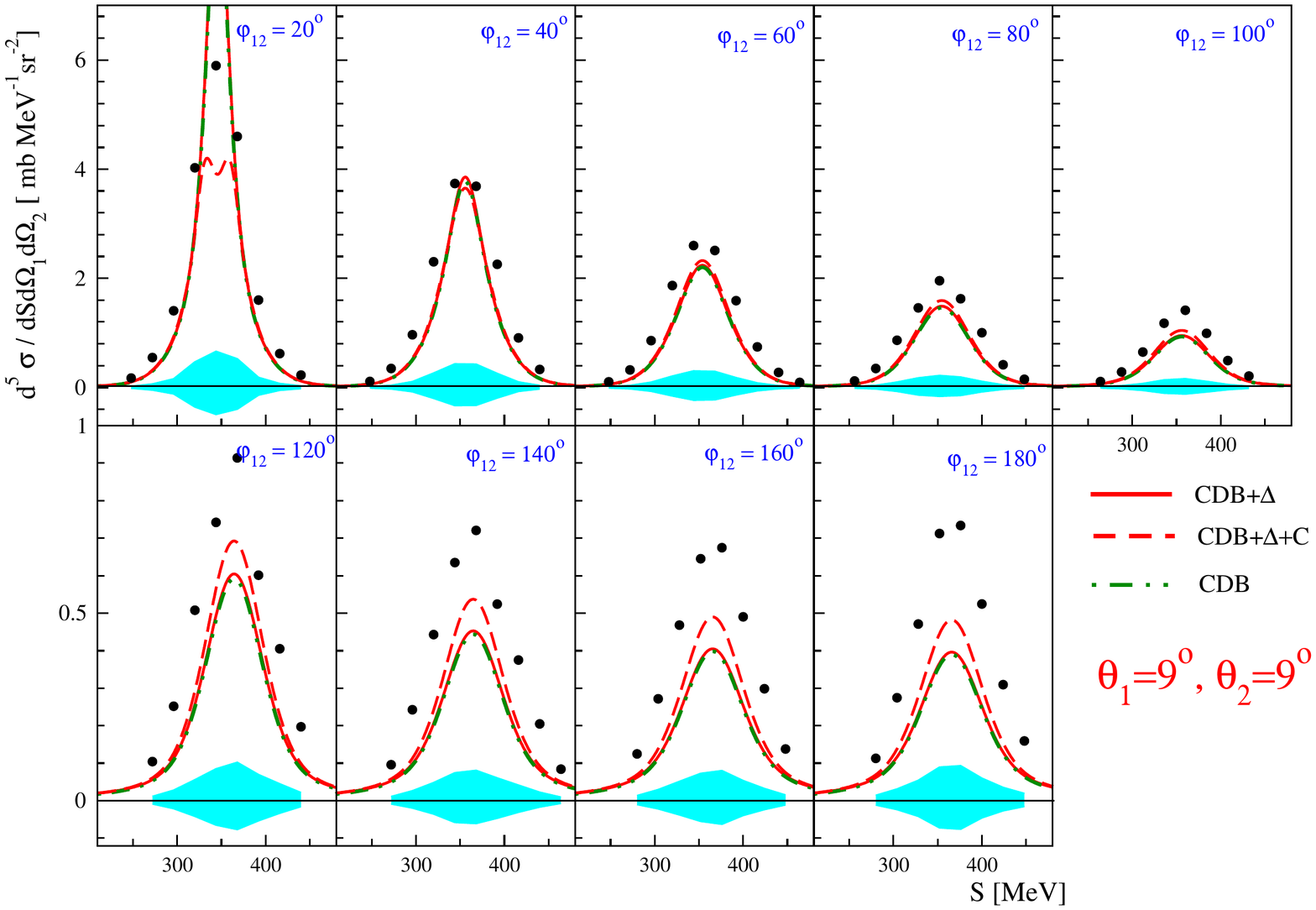}
}
\caption{(Color online) The same as Fig.~\ref{fig5t5} but for kinematic configurations with the proton polar angles  $\theta _1$=7$^\circ$, $\theta _2$=15$^\circ$ and $\theta _1$=9$^\circ$, $\theta _2$=9$^\circ$. 
\label{fig7t15}}
\end{figure*}
\end{center}  

\begin{center}
\begin{figure*}[h]
\subfigure{
\includegraphics[width=16.5cm]{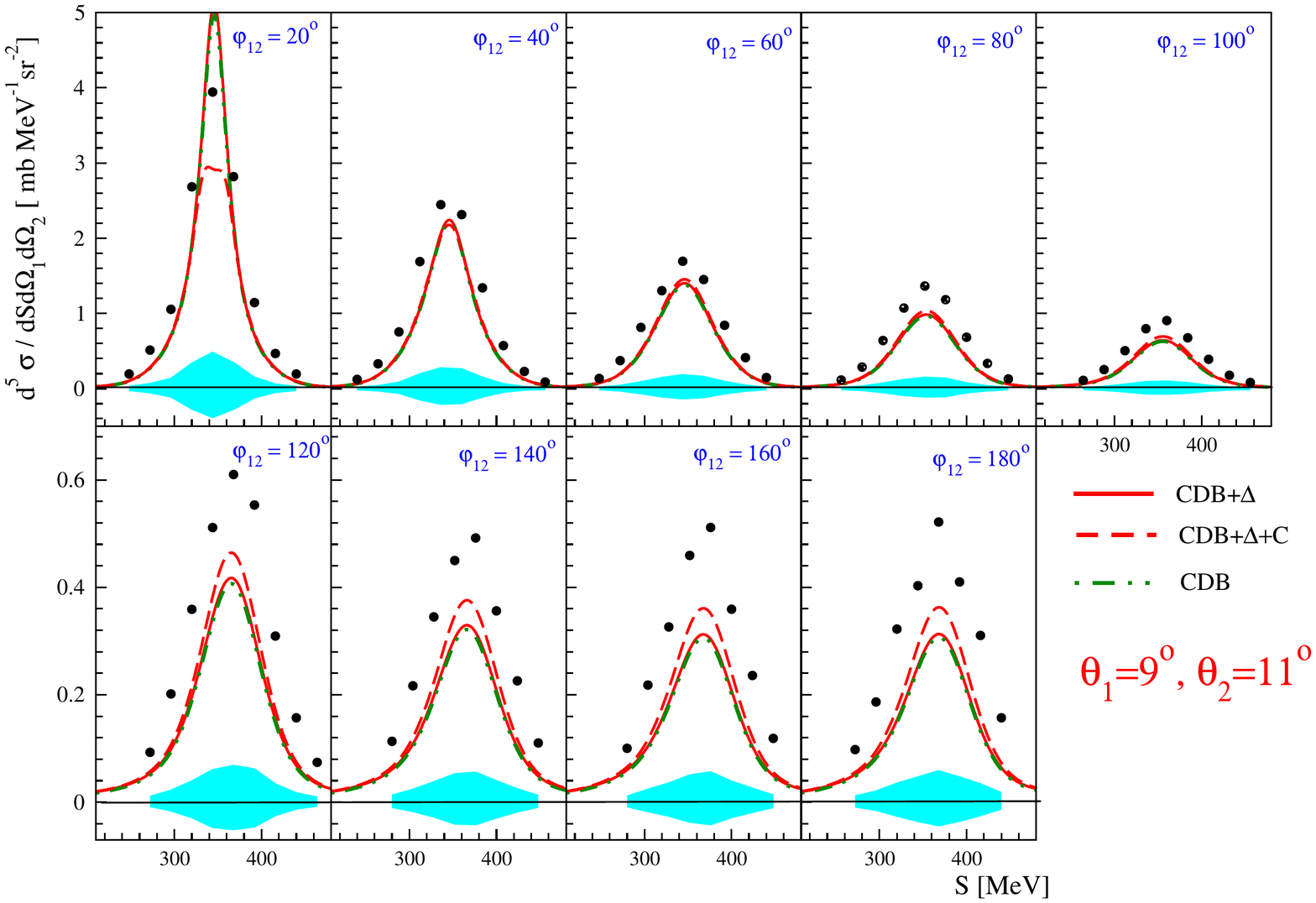}
}
\subfigure{
\includegraphics[width=16.5cm]{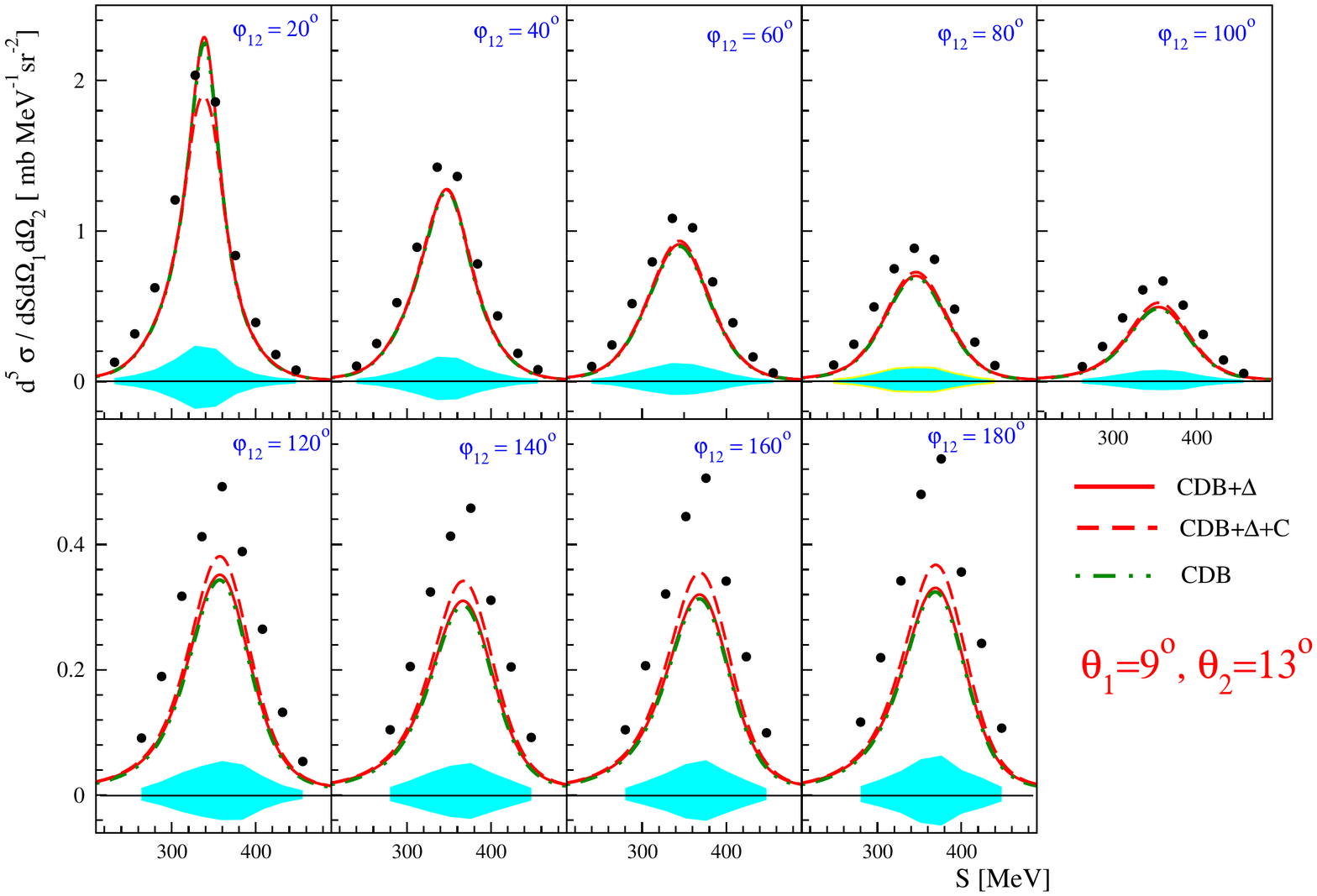}
}
\caption{(Color online) The same as Fig.~\ref{fig5t5} but for kinematic configurations with the proton polar angles  $\theta _1$=9$^\circ$, $\theta _2$=11$^\circ$ and $\theta _1$=9$^\circ$, $\theta _2$=13$^\circ$. 
\label{fig9t11}}
\end{figure*}
\end{center}  

\begin{center}
\begin{figure*}[h]
\subfigure{
\includegraphics[width=16.5cm]{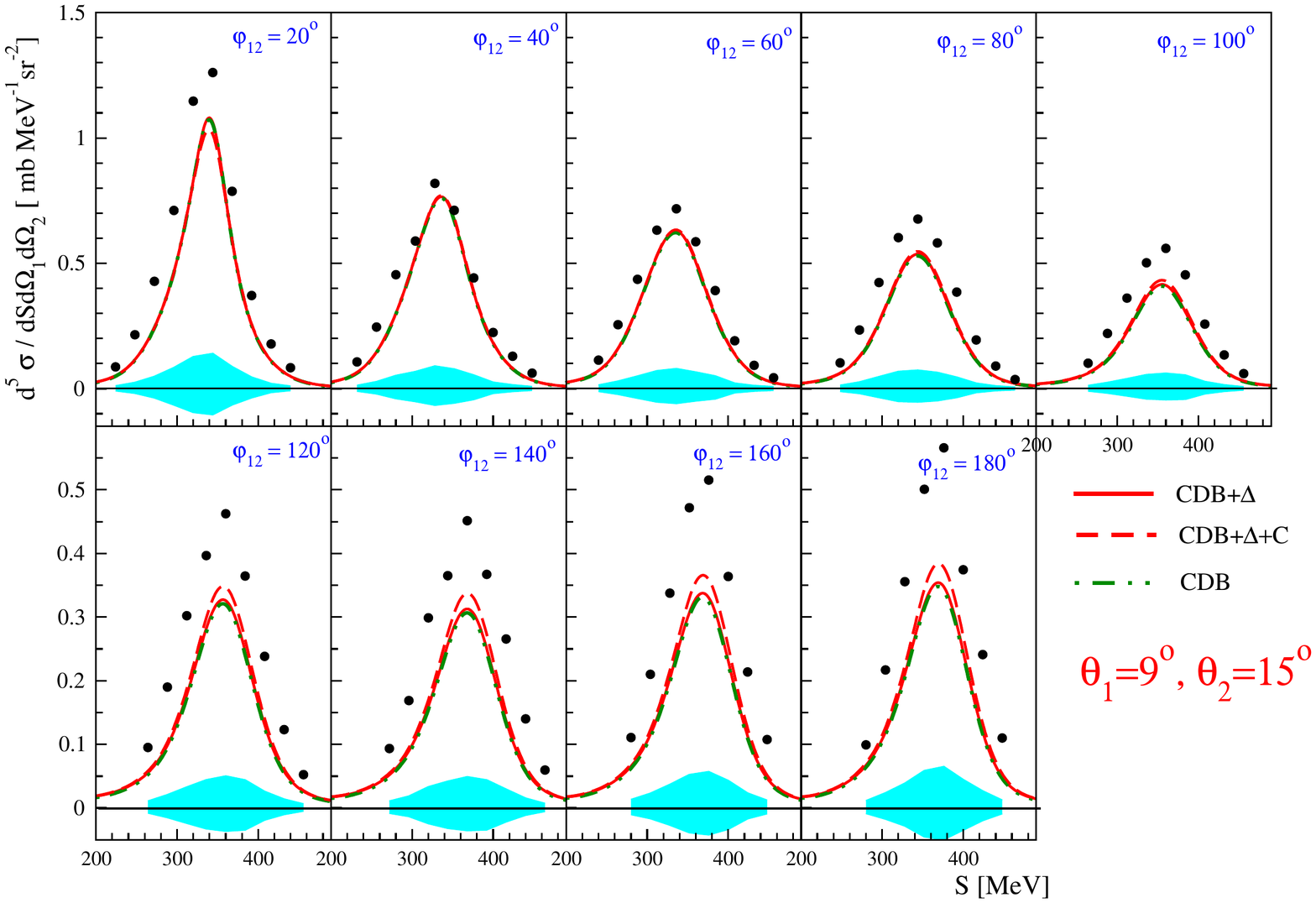}
}
\subfigure{
\includegraphics[width=16.5cm]{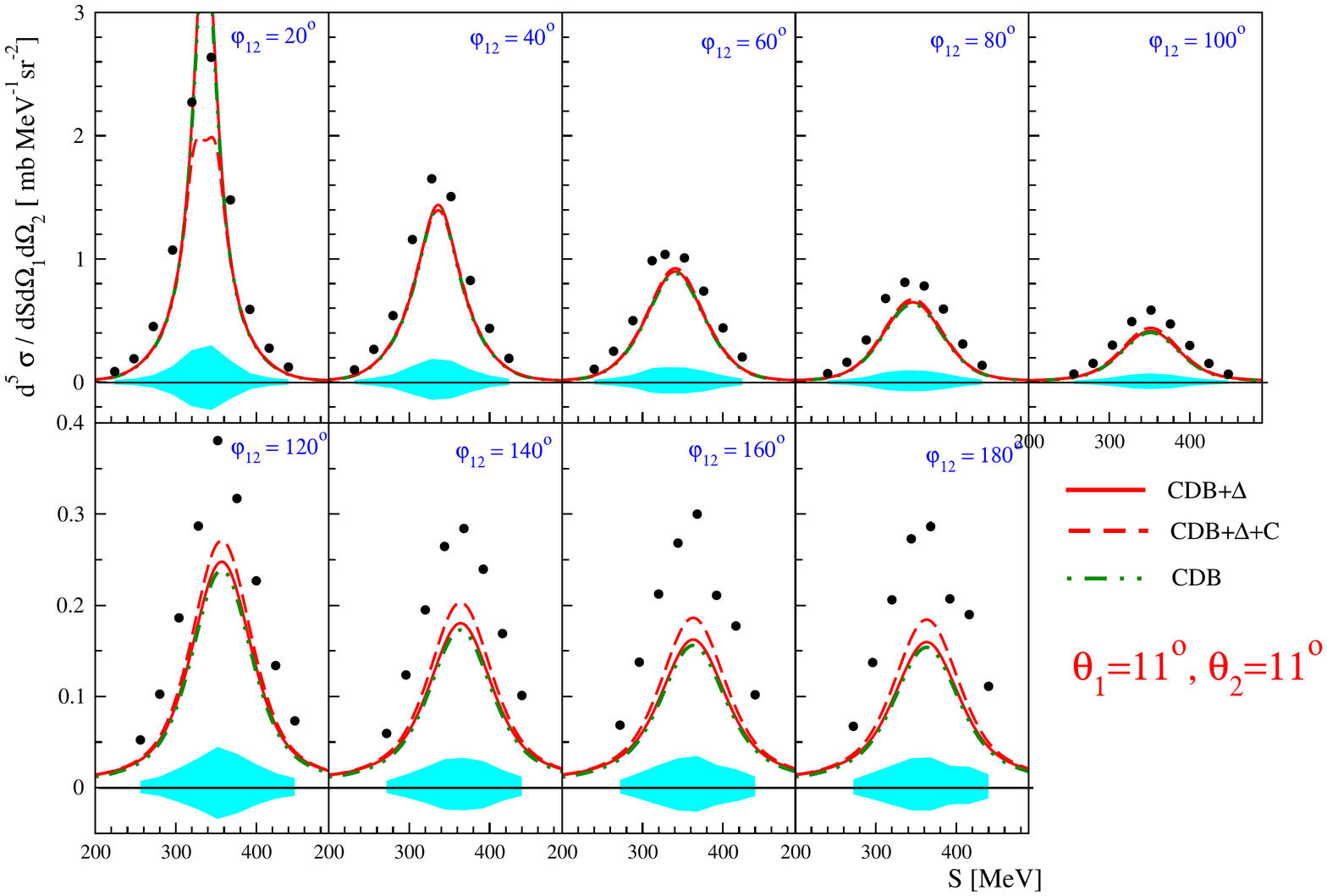}
}
\caption{(Color online) The same as Fig.~\ref{fig5t5} but for kinematic configurations with the proton polar angles  $\theta _1$=9$^\circ$, $\theta _2$=15$^\circ$ and $\theta _1$=11$^\circ$, $\theta _2$=11$^\circ$. 
\label{fig9t15}}
\end{figure*}
\end{center}  

\begin{center}
\begin{figure*}[h]
\subfigure{
\includegraphics[width=16.5cm]{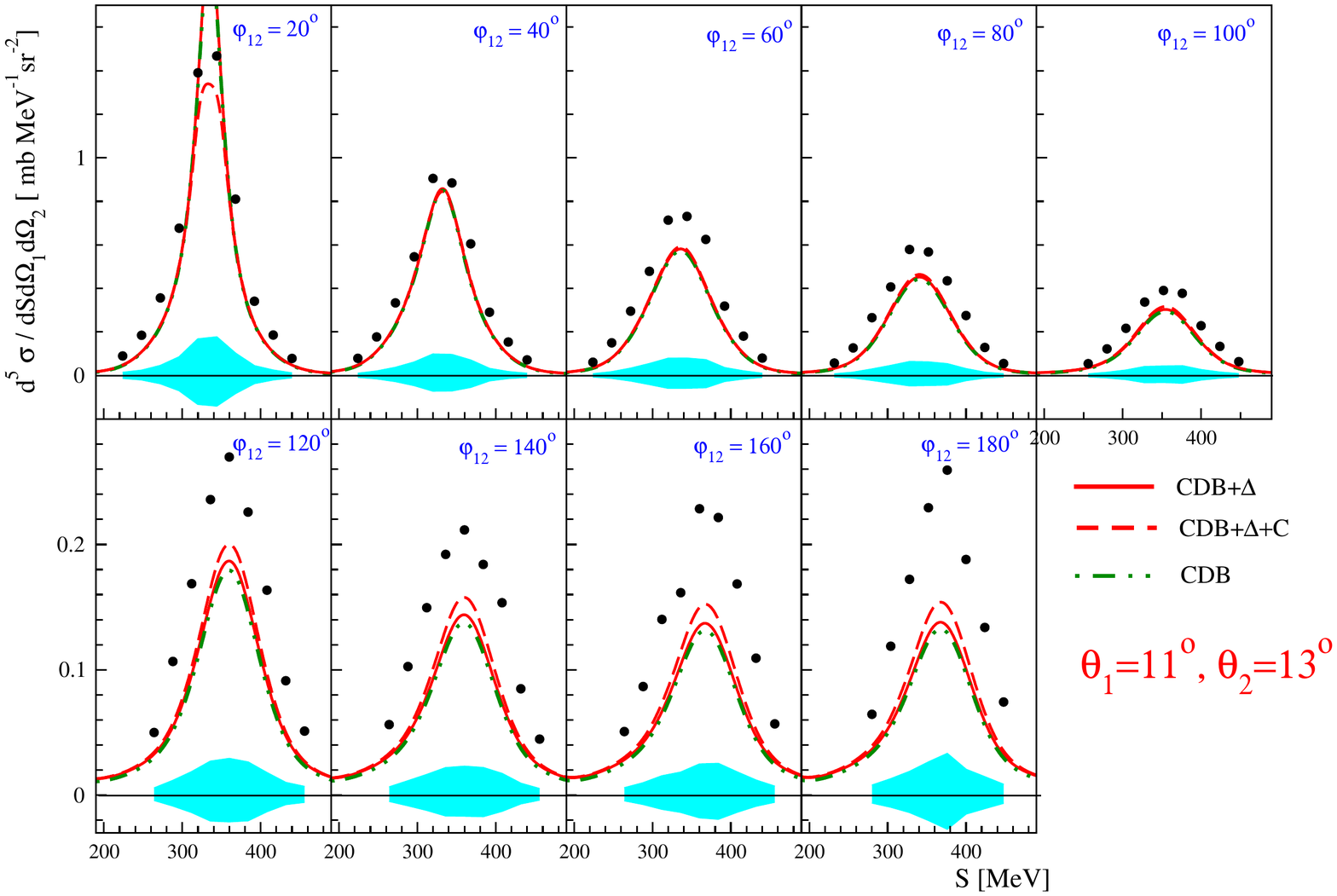}
}
\subfigure{
\includegraphics[width=16.5cm]{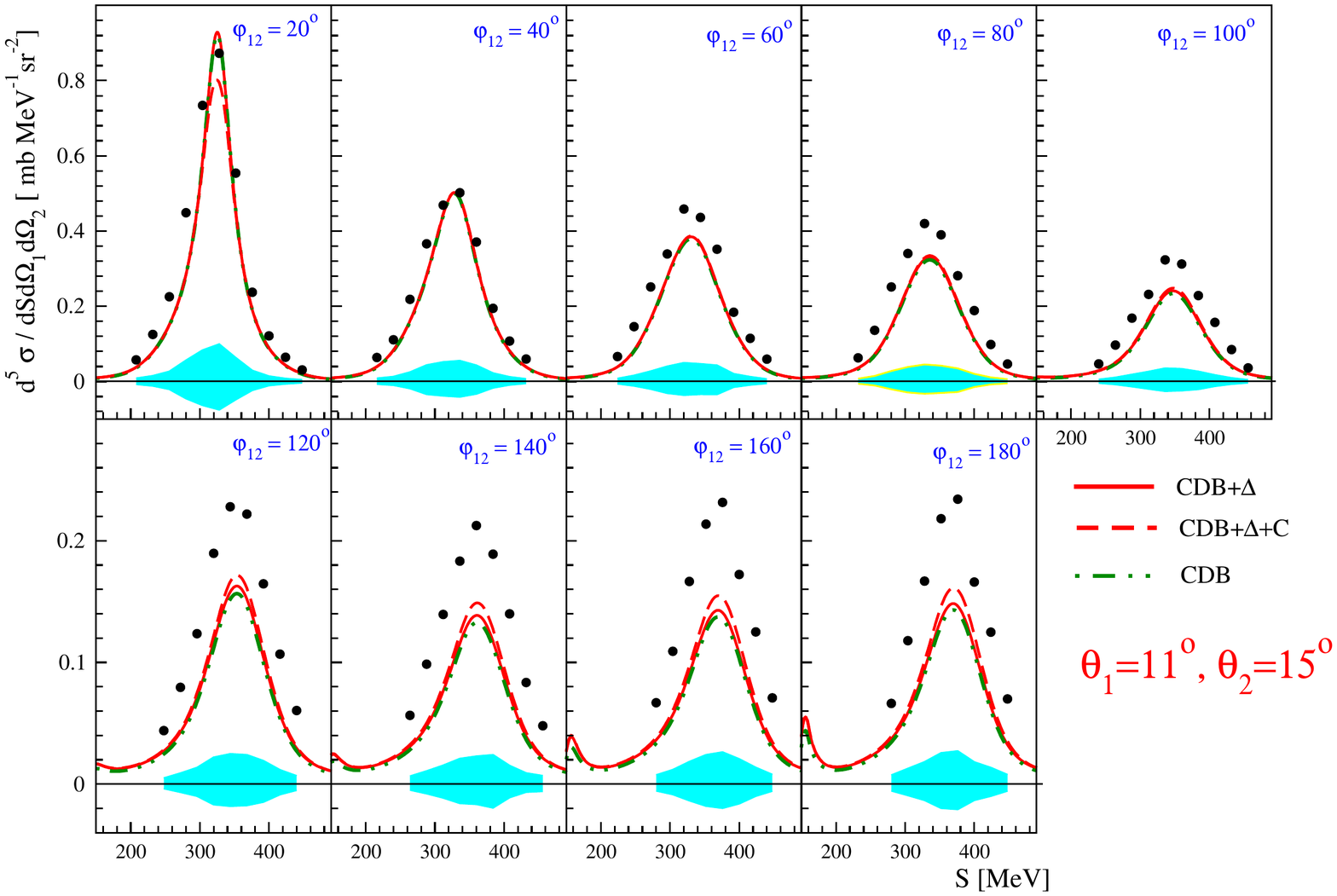}
}
\caption{(Color online) The same as Fig.~\ref{fig5t5} but for kinematic configurations with the proton polar angles  $\theta _1$=11$^\circ$, $\theta _2$=13$^\circ$ and $\theta _1$=11$^\circ$, $\theta _2$=15$^\circ$. 
\label{fig11t13}}
\end{figure*}
\end{center}  

\begin{center}
\begin{figure*}[h]
\subfigure{
\includegraphics[width=16.5cm]{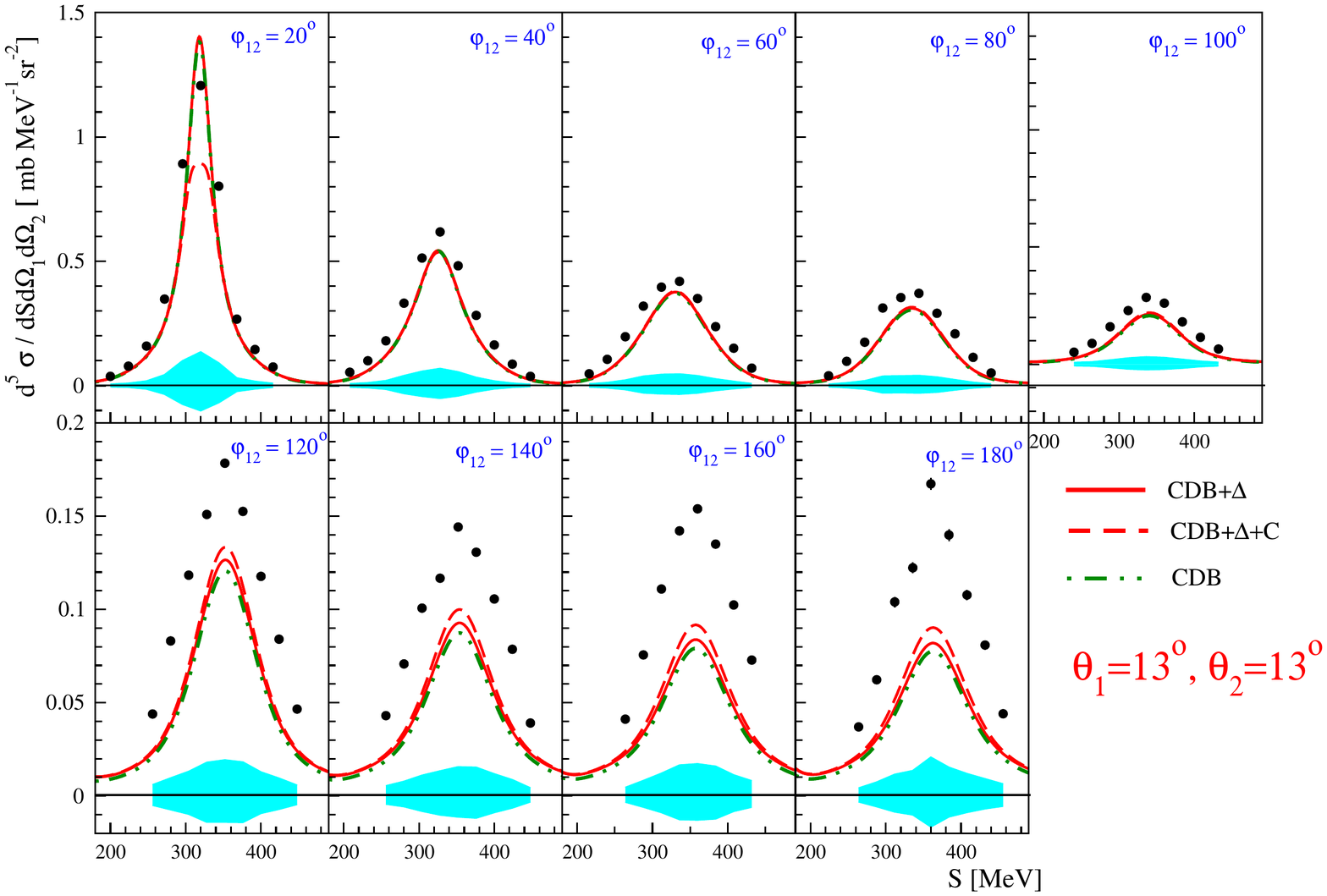}
}
\subfigure{
\includegraphics[width=16.5cm]{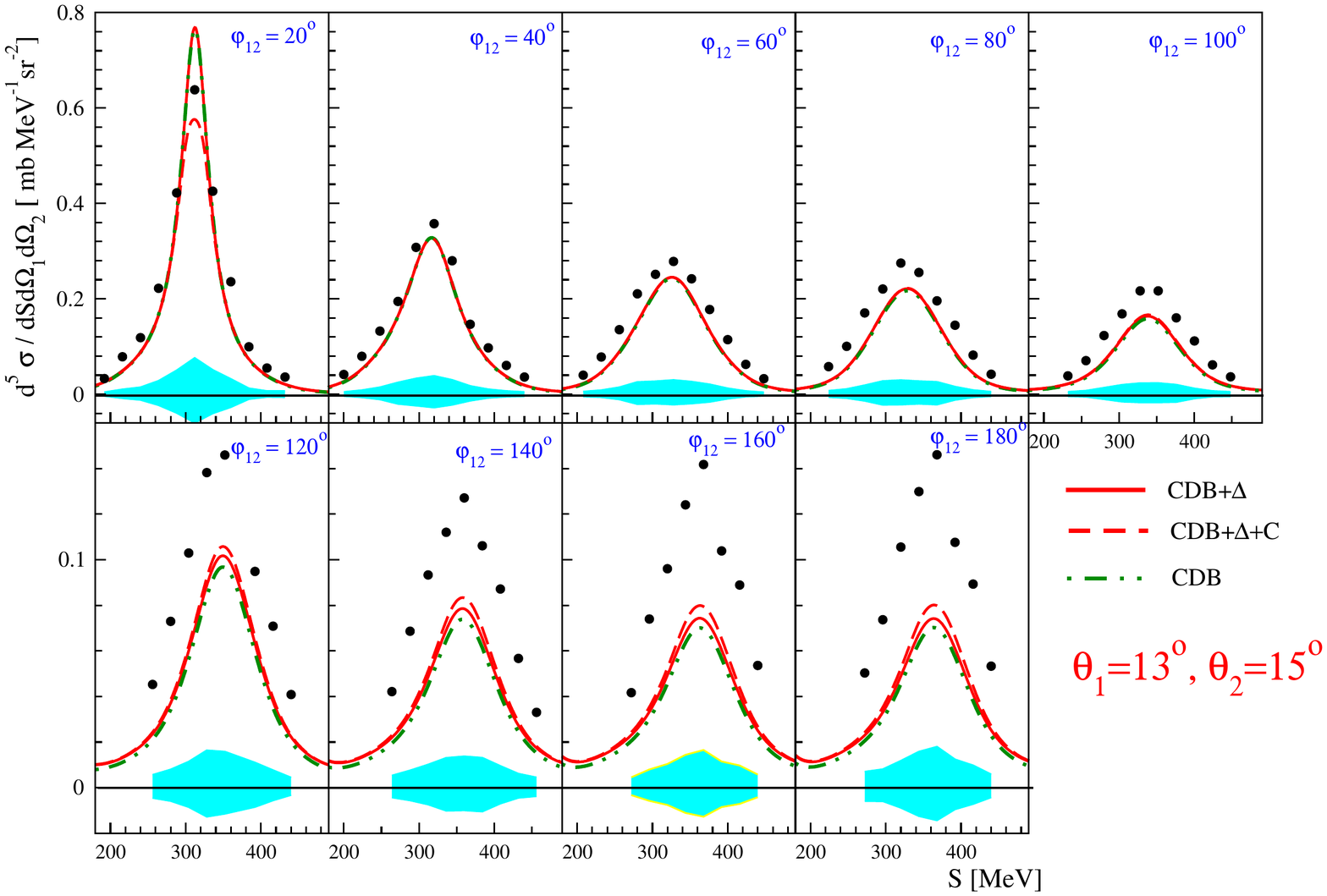}
}
\caption{(Color online) The same as Fig.~\ref{fig5t5} but for kinematic configurations with the proton polar angles  $\theta _1$=13$^\circ$, $\theta _2$=13$^\circ$ and $\theta _1$=13$^\circ$, $\theta _2$=15$^\circ$. 
\label{fig13t13}}
\end{figure*}
\end{center}  

\begin{center}
\begin{figure*}[h]

\includegraphics[width=16.5cm]{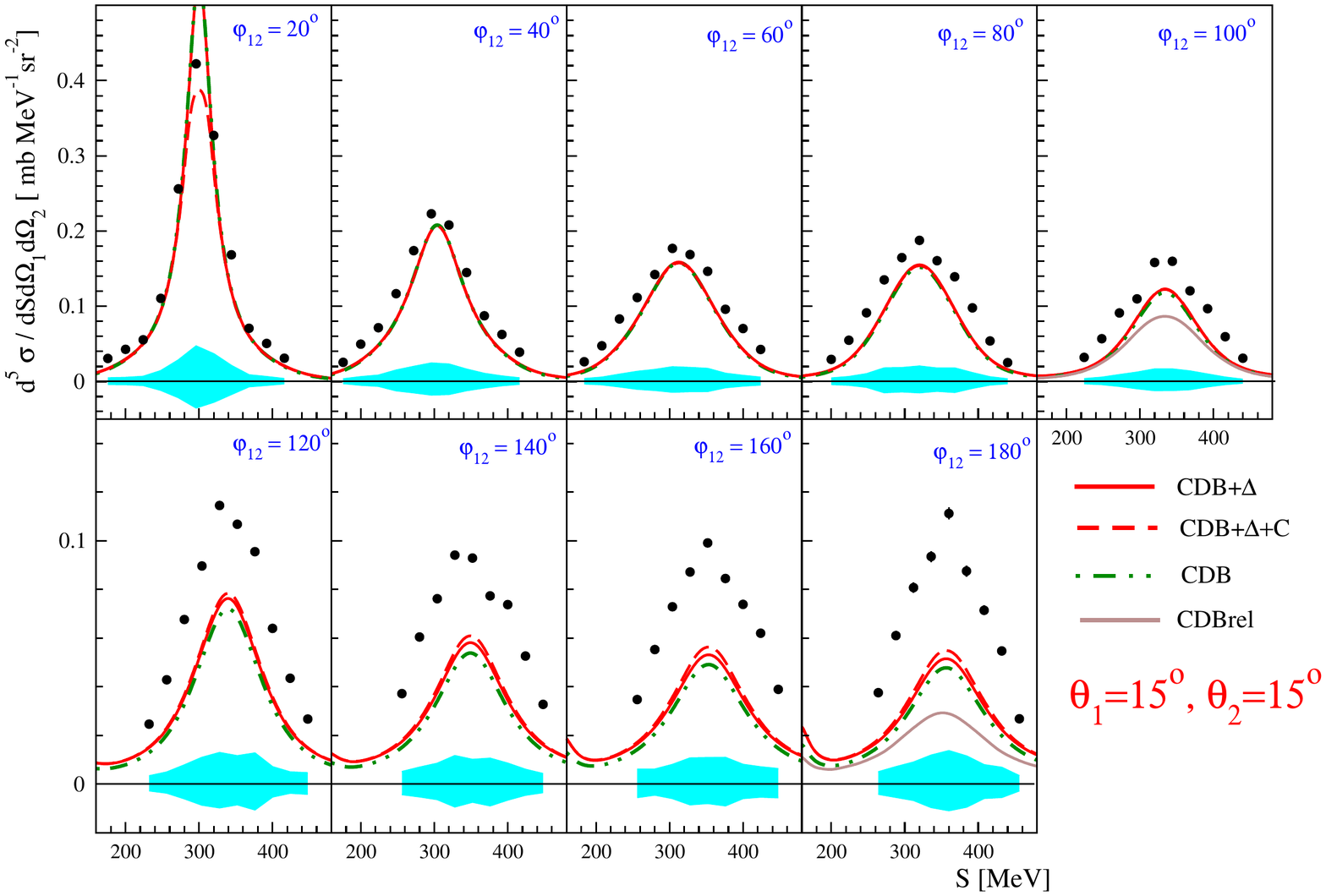}

\caption{(Color online) The same as Fig.~\ref{fig5t5} but for kinematic configurations with proton polar angles  $\theta _1$=15$^\circ$, $\theta _2$=15$^\circ$ . 
\label{fig15t15}}
\end{figure*}
\end{center}  

\bibliography{bibliography1}
\end{document}